\documentclass[parskip=half,american]{scrartcl}
\usepackage[T1]{fontenc}
\usepackage{textcomp}
\usepackage[utf8]{inputenc}
\usepackage{xcolor}
\usepackage{babel}
\usepackage{booktabs}
\usepackage{url}
\usepackage{varwidth}
\usepackage{amsmath}
\usepackage{amssymb}
\usepackage{graphicx}
\usepackage{geometry}
\PassOptionsToPackage{normalem}{ulem}
\usepackage{ulem}
\usepackage[pdfusetitle,
 bookmarks=true,bookmarksnumbered=true,bookmarksopen=false,
 breaklinks=false,pdfborder={0 0 1},backref=false,colorlinks=false]
 {hyperref}

\makeatletter

\providecommand{\tabularnewline}{\\}
\newenvironment{cellvarwidth}[1][t]
    {\begin{varwidth}[#1]{\linewidth}}
    {\@finalstrut\@arstrutbox\end{varwidth}}
\providecolor{lyxadded}{rgb}{0,0,1}
\providecolor{lyxdeleted}{rgb}{1,0,0}
\DeclareRobustCommand{\mklyxadded}[1]{\textcolor{lyxadded}\bgroup#1\egroup}
\DeclareRobustCommand{\mklyxdeleted}[1]{\textcolor{lyxdeleted}\bgroup\mklyxsout{#1}\egroup}
\DeclareRobustCommand{\mklyxsout}[1]{\ifx\\#1\else\sout{#1}\fi}

\newcommand{\lyxaddress}[1]{
	\par {\raggedright #1
	\vspace{1.4em}
	\noindent\par}
}

\usepackage{pifont}
\newcommand{\xmark}{\text{\ding{55}}}

\makeatother

\begin{document}
\title{Review of color difference measures}
\author{Patrick De Visschere}
\maketitle

\lyxaddress{Ghent University, Department Electronics and Information Systems,
Liquid Crystals and Photonics.\\
Technologiepark-Zwijnaarde 126, BE-9052 Gent, email: PatrickDeVisschere@UGent.be,
ORCID: \url{https://orcid.org/0000-0003-0278-8199}}
\begin{abstract}
We made a detailed review of the difference measures which have been
used to judge the differences between experimentally determined color
differences and theoretically defined ones, so-called line elements,
for the human visual system. To eliminate the statistical errors due
to variable and usually arbitrary sampling of the directions in a
color point, we integrate the measures over a complete ellipsoid/ellipse.
It turns out that in the limit for small deviations from circularity
all proposed measures ($V_{AB}$, $\gamma-1$, $CV$ and $\mathrm{STRESS}$)
are equivalent. For greater deviations the measures become distinct
with $\gamma-1$ the most sensitive and $\mathrm{STRESS}$ the least.
Ideally a difference measure should be coordinate independent and
then it is advantageous to apply an affine transformation to both
sets, e.g. turning the theoretical one into the unit ball. Although
MacAdam already used this method but sampled the transformed ellipse,
we integrate over the ellipsoid/ellipse. Comparing the results with
the base measures we show that only $\mathrm{STRESS}$ is coordinate
independent. Judging whether a single ellipsoid/ellipse resembles
a unit ball can easily be done by comparing the eigenvalues with one
and we show that our previously proposed error measure $d_{ev}$ (Candry
e.a. Optics Express, 30, 36307, 2022) is the eigenvalue version of
$\gamma-1$. We show that the correlation coefficient $r$ although
being hypersensitive to orientation mismatches depends on the orientation
coordinate even if it has no physical meaning, a clear manifestation
of its coordinate dependence, but that Pant's recent geometric measure
$1-R$ on the other hand is coordinate independent. All measures are
routinely made scale invariant by the introduction of a scaling parameter,
to be optimized. Lastly we show that from all measures the $\gamma-1$
ones are the only ones permitting a simple derivation of the globally
optimized difference measure from the locally defined ones.
\end{abstract}

\section{Introduction.}

It is well established now that for the Human Visual System (HVS)
no color coordinates are known in which the perceptual difference
between two colors is given by their Euclidean distance. This is obviously
not so for the CIE $(X,Y,Z)$ coordinates but also the approximately
uniform color spaces $\left(L^{*},a^{*},b^{*}\right)$ or $\left(L^{\prime},u^{\prime},v^{\prime}\right)$
fall short of this goal. Following these initial proposals, many different,
steadily improving, color difference expressions have been presented
culminating in the reasonably uniform $\mathrm{CIEDE2000}$ expression
\cite{Luo:2001hd}. Although no explicit reference is made to it,
the color space underlying this model is a Riemann space, which is
only Euclidean over infinitesimal distances and the metric changes
from point to point. It is also known that the HVS is much more complicated
than initially thought and a more complete description is given by
Color Appearance Models (CAMs), but at the heart of these CAMs one
will invariably find a color difference model.

The standard way to define a Riemann space is to define the metric
tensor as a function of suitable coordinates $x^{i}$. This tensor
defines a so called line element (LE), the infinitesimal distance
$d\sigma$ between two nearby color points with coordinates $x^{i}$
and $x^{i}+dx^{i}$
\begin{equation}
d\sigma^{2}=g_{ij}\left(x\right)dx^{i}dx^{j}\label{eq:generic LE}
\end{equation}

To find the distance between two color points $A$ and $B$ one must
then integrate this LE along the geodesic between those points, that
is along the path with the extremal distance
\begin{equation}
\sigma=\int^{B}_{A}d\sigma\label{eq:global color difference}
\end{equation}

but it is known \cite{MacAdam:1963ub} that the HVS is strictly speaking
not Riemannian on this global scale and this distance thus does not
directly match well with the perceived difference. It is important
to notice that although superficially a tensor looks like a matrix,
a tensor has an intrinsic physical meaning independent of the coordinates.
But in practice a tensor $g$ is represented by its matrix elements
$g_{ij}$ in a particular coordinate system $x^{i}$ and the use of
a particular coordinate system is unavoidable. The line element $d\sigma$
in eq.(\ref{eq:generic LE}) and the color difference between two
color points measured along a geodesic in eq.(\ref{eq:global color difference})
are such coordinate independent properties. Since color difference
measures in reality compare 2 tensors (an experimental one and a model
one) it seems then natural to prefer difference measures which are
also coordinate independent.

In order to judge those difference equations a plethora of difference
measures has been introduced. With these a comparison can be made
between observed (experimental) differences, usually denoted by $\Delta V_{k}$
and calculated (modeled) differences, usually denoted by $\Delta E_{k}$.
Both these quantities are of the type $d\tau$ as in eq.(\ref{eq:generic LE})
but belong to different metric tensors. Given a set $\left\{ \Delta V_{k}\right\} $
and a corresponding set $\left\{ \Delta E_{k}\right\} $ a particular
difference measure gives an estimate of how well these sets match
each other, with a perfect match resulting in a zero difference. An
overview of these measures has been given by Luo \cite{Luo:2002et}
and Garcia \cite{Garcia:2007fz}.

The $V_{AB}$ measure is defined by
\begin{equation}
V^{2}_{AB}=\left\langle \frac{\left(\Delta E_{k}-F\Delta V_{k}\right)^{2}}{\Delta E_{k}F\Delta V_{k}}\right\rangle \label{eq:VABbase}
\end{equation}

where $\left\langle .\right\rangle $ denotes averaging over the set
(For the time being we consider a set of $(\Delta V,\Delta E)$ pairs
at or near a single point in color space. The extension to a set of
such points will be made later.) and where $F$ is a scaling factor.
We will always choose such a scaling factor optimally, in this case
meaning that
\begin{equation}
F_{opt}=\sqrt{\frac{\left\langle \mathrm{\left(E/V\right)}_{k}\right\rangle }{\left\langle \mathrm{\left(E/V\right)^{-1}_{k}}\right\rangle }}\label{eq:VABFopt}
\end{equation}

where we defined $\mathrm{{E/V}=\frac{\Delta E}{\Delta V}}$, and
resulting in the optimally scaled expression
\begin{equation}
V^{2}_{AB,opt}=2\left(\sqrt{\left\langle \mathrm{\left(E/V\right)}_{k}\right\rangle \left\langle \mathrm{\left(E/V\right)^{-1}_{k}}\right\rangle }-1\right)\label{eq:VAB}
\end{equation}

This (optimal) measure is then \emph{scale invariant} and obviously
also \emph{symmetric}, that is invariant for the exchange of $\Delta V$
and $\Delta E$ and the order of these variables is thus of no concern.
All measures considered include a scale factor $F$, which allows
to scale one ellipsoid w.r.t. the other. If the ellipsoids to be compared
are merely scaled versions of each other (and properly aligned) then
a perfect match will be obtained. One should handle this scale factor
properly: in a typical experiment one considers different color points
and although a different scale factor can be defined in each of these
color point, when evaluating the dataset one should define a single
scale factor for the set.

The $\gamma$ measure is defined by (The original definition used
a base 10 logarithm, but a natural logarithm seems more natural.)
\begin{equation}
\left(\ln\gamma\right)^{2}=\left\langle \left(\ln F\mathrm{\left(E/V\right)}_{k}\right)^{2}\right\rangle \label{eq:gammabase}
\end{equation}

and with the optimal scaling factor given by
\begin{equation}
\ln F_{opt}=-\left\langle \ln\left(E/V\right)_{k}\right\rangle \label{eq:gammaFopt}
\end{equation}

we get the optimal result
\begin{equation}
\left(\ln\gamma_{opt}\right)^{2}=\left\langle \left(\ln\mathrm{\left(E/V\right)}_{k}\right)^{2}\right\rangle -\left\langle \ln\mathrm{\left(E/V\right)}_{k}\right\rangle ^{2}\label{eq:gamma}
\end{equation}

and is also \emph{symmetric}.

Finally $\mathrm{CV}$ is defined by
\begin{equation}
\mathrm{CV}^{2}=\frac{\left\langle \left(\Delta E_{k}-F\Delta V_{k}\right)^{2}\right\rangle }{\left\langle \Delta E_{k}\right\rangle ^{2}}\label{eq:CVbase}
\end{equation}

with the optimal scaling factor 
\begin{equation}
F_{opt}=\frac{\left\langle \Delta E_{k}\Delta V_{k}\right\rangle }{\left\langle \Delta V^{2}_{k}\right\rangle }\label{eq:CVFopt}
\end{equation}
and optimal result given by
\begin{equation}
CV^{2}_{opt}=\frac{\left\langle \Delta E^{2}_{k}\right\rangle \left\langle \Delta V^{2}_{k}\right\rangle -\left\langle \Delta E_{k}\Delta V_{k}\right\rangle ^{2}}{\left\langle \Delta E_{k}\right\rangle ^{2}\left\langle \Delta V^{2}_{k}\right\rangle }\label{eq:CV}
\end{equation}

where we note the \emph{asymmetry} in the denominator although $CV$
is obviously now \emph{scale invariant}. Measures $CV$ (Coates defined
$\mathrm{CV}$ as a percent, but in line with the other metrics, we
just define it as a fraction.) and $\gamma$ were introduced by Coates
\cite{Coates:1981lt} and $V_{\mathrm{AB}}$ by Schultze \cite{Schultze:1972sm}.
A fourth \emph{symmetric}, \emph{scale invariant} measure, a conventional
correlation coefficient $r$, has also been considered
\begin{equation}
r=\frac{\left\langle \Delta E_{k}\Delta V_{k}\right\rangle -\left\langle \Delta E_{k}\right\rangle \left\langle \Delta V_{k}\right\rangle }{\sqrt{\left(\left\langle \Delta E^{2}_{k}\right\rangle -\left\langle \Delta E_{k}\right\rangle ^{2}\right)\left(\left\langle \Delta V^{2}_{k}\right\rangle -\left\langle \Delta V_{k}\right\rangle ^{2}\right)}}\label{eq:r}
\end{equation}

but according to Luo ``...in some cases the correlation coefficient
was quite inconsistent with the other measures ...'' \cite{Luo:2002et}.
We note that (\ref{eq:r}) is known as the Pearson correlation coefficient
in statistics and takes values in the interval $\left[-1,1\right]$.
As such it is not usable as a difference measure. For 2 perfectly
correlated variables ($\Delta V=\Delta E$) one finds of course $r=1$,
but variables can also be anticorrelated ($r=-1$). The least one
can do is to consider e.g. $\bar{r}=1-r$ as a more proper difference
measure which then takes values in the interval $\left[0,2\right]$
and with perfectly anti correlated variables yielding then $\bar{r}=2$.
In two recent papers \cite{Kirchner:2011zy}\cite{Carter:2022nr}
a comparison was made between this correlation coefficient $r$ and
the STRESS measure, to be discussed below, with Kirchner e.a. being
in favor of the former measure. Carter e.a. advise the use of centered
data (that is wit a zero avearge) and shows that in that case both
measures are equivalent. We will consider these contradicting evaluations
in §~\ref{subsec:The-correlation-coefficient.}.

A popular difference measure $\mathrm{PF}/3$ averages the three previous
measures $V_{AB}$, $\bar{\gamma}=\gamma-1$ and $CV$ \cite{Luo:1987ji,Guan:1999es}
\begin{equation}
\mathrm{PF/3}=\frac{1}{3}\left(V_{AB}+\bar{\gamma}+CV\right)\label{eq:PF3}
\end{equation}

The reason for taking this average looks somewhat arbitrary and Guan
\cite{Guan:1999es} merely notices that ``The PF also eases the comparison,
since one value is obtained, and avoids making a decision as to which
of the measures is the best.''

More recently another difference measure has become popular, called
the $\mathrm{STRESS}$ measure, standing for ``Standardized Residual
Sum of Squares,'' and finding its origin in ``Multidimensional Scaling''
\cite{Kruskal:1964ap}
\begin{equation}
\mathrm{{STRESS}}^{2}=\frac{\left\langle \left(F\Delta E_{k}-\Delta V_{k}\right)^{2}\right\rangle }{\left\langle \Delta V^{2}_{k}\right\rangle }\label{eq:STRESSbase}
\end{equation}

with again a scale factor $F$, which can be chosen optimally as
\begin{equation}
F_{\mathrm{opt}}=\frac{\left\langle \Delta E_{k}\Delta V_{k}\right\rangle }{\left\langle \Delta E^{2}_{k}\right\rangle }\label{eq:STRESSFopt}
\end{equation}

and with the optimum (There are a number of alternative definitions
for (\ref{eq:STRESSbase}) but since they all lead to the same optimal
result, we only mention one of them. For the others see \cite{Garcia:2007fz}.)
\begin{equation}
\mathrm{{STRESS}^{2}_{opt}}=1-\frac{\left\langle \Delta V_{k}\Delta E_{k}\right\rangle ^{2}}{\left\langle \Delta E^{2}_{k}\right\rangle \left\langle \Delta V^{2}_{k}\right\rangle }\label{eq:STRESS}
\end{equation}

STRESS is thus again \emph{scale invariant} and also \emph{symmetric}.
A related metric which permits for significance F-testing is $V_{M}$
\cite{Garcia:2007fz} 
\begin{equation}
V_{M}=\frac{N}{N-1}\left\langle \left(F\Delta E_{k}-\Delta V_{k}\right)^{2}\right\rangle 
\end{equation}

where the scale factor $F$ has the same optimal value as in (\ref{eq:STRESSFopt})
and with optimal result
\begin{equation}
V_{M,\mathrm{opt}}=\frac{N}{N-1}\left\langle \Delta V^{2}_{k}\right\rangle \left[1-\frac{\left\langle \Delta V_{k}\Delta E_{k}\right\rangle ^{2}}{\left\langle \Delta E^{2}_{k}\right\rangle \left\langle \Delta V^{2}_{k}\right\rangle }\right]
\end{equation}

This specific measure allows to decide whether 2 different models
($\Delta E_{A}$ and $\Delta E_{B}$) are significantly different
w.r.t. experimental data $\Delta V$. Due to the additional average
$V_{M}$ is \emph{not symmetric} and \emph{not scale invariant}. It
can be considered as a specific application of $\mathrm{STRESS}$
and we will no longer consider it separately.

In a recent paper \cite{Candry:2022kx}, where we proposed a new LE,
further on denoted by RieLE (Being a Riemann LE but also reminiscent
of the LE's proposed by Friele, on which it is loosely based.), we
also proposed a novel method to compare differences, denoted here
as $d_{ev}$. All difference measures mentioned so far are \emph{direct}
measures, which compare $\Delta V_{k}$ and $\Delta E_{k}$ obtained
in one or more color points and in several directions through those
points. It is obvious that to define the metric tensor in (\ref{eq:generic LE})
properly one must sample those directions as uniformly and as densely
as possible. This seems to be the weak point of the previously defined
difference measures: the great variability in the selection of sample
directions. Whereas in his famous experiments, limited to chromaticity
variations, MacAdam considered 5-9 directions in each color point,
with an average of $171/25=6.8$, which seemed mostly adequate, in
more recent experiments sampling is often limited, e.g. Alman in their
(chromaticity) phase I experiment \cite{Alman:1989zo} considered
only 4 directions and in their full phase II experiment \cite{Berns:1991ij}
this was enlarged to 6, still smaller than Macadam's 2-dimensional
average. The situation is further confounded by the distribution of
$\Delta V$ values, needed to define a proper psychometric function.
Actually we are comparing 2 families of scaled ellipsoids/ellipses.

In our difference measure we compare the ellipsoids/ellipses with
each other instead of directly comparing the color differences $\Delta V_{k}$
and $\Delta E_{k}$. We thus favor a two step process: on the basis
of the measurements $\Delta V_{k}$ the parameters of the metric tensor
$g_{ij}$ are defined and here also properly sampling of all directions
and sizes is paramount to obtain an adequate description. In a 2nd
and separate step the tensor found is compared with a theoretical
tensor, defining a line element. This comparison can be done based
on the eigenvalues and eigenvectors of the ellipsoids and there is
no need to sample them, resulting in a much simpler difference measure.
Since a metric tensor is completely defined by its eigenvalues and
its orthonormal eigenvectors 6 parameters (3 eigenvalues, 2 components
of one eigenvector and 1 rotation angle around this vector) at most
are needed to define a full metric tensor (3 for a chromaticity tensor).
But for comparing two ellipsoids initially only 9 (or 5) parameters
are needed, because only the relative orientation does matter. Eventually,
taking the scaling into account, only 2 (or 1) remain as will become
clear. All directly defined difference measures can in principle also
be defined using eigenvalues and it turns out that our newly defined
difference measure is related in this sense to $\bar{\gamma}$.

We believe this 2-step model was also implemented by MacAdam \cite{MacAdam:1964hk},
but where in the 2nd step also a discrete sampling was applied (over
the unit circle). Since we want to relegate all statistical issues
to the first step we avoid all sampling in step 2 and to compare our
(eigenvalue) based method with the direct method, in the latter case,
instead of summing over some samples we integrate over the unit circle.
These continuous measures allow to compare the different difference
metrics on an equal footing an without effects of inadequate sampling.

The formulation given so far is adequate when considering a single
color point and all discrete averages in eq.(\ref{eq:VABcont})-(\ref{eq:CVcont})
are then taken over the directions sampled in that color point. When
considering a complete data set encompassing a representative set
of color points we index the distances by two indices, e.g. $\Delta V_{kl}$
where $k$ indexes the color points and $l$ the directions considered
at point $k$ and the averages are then taken over all directions
in all color points. It is important that in this latter case only
a single optimal scale factor $F_{\mathrm{opt}}$ is chosen for the
complete data set, since all observations are supposed to have been
made under similar circumstances. The result being exactly the same
as for a single color point using double indices is not really necessary.
To obtain the (globally optimized) difference in one of the color
points one must then apply the base definitions (\ref{eq:VABbase})(\ref{eq:gammabase})(\ref{eq:CVbase})(\ref{eq:STRESSbase})
but with $F=F_{\mathrm{opt}}$ the globally optimized scale factor.
In addition one can still consider a local optimization in each color
point to be calculated with the optimized eqs.(\ref{eq:VAB})(\ref{eq:gamma})(\ref{eq:CV})(\ref{eq:STRESS})
and these locally optimal values will obviously be smaller than those
obtained with the global $F_{\mathrm{opt}}$. It would be interesting
if the former (globally optimized) local measures could be found from
the locally optimized ones and this is indeed feasible for the $\bar{\gamma}$
measure (and our $d_{ev}$ which is very similar). For the other measures
it is sometimes also possible but more complicated and requires the
use of weighted averages.

Finally an interesting but not widely known and even less used difference
measure was introduced by Pant \cite{Raj-Pant:2012kb}. The measure
is also defined by comparing ellipses (The method could equally well
be applied to ellipsoids but as far as we known this has not been
done.) but purely geometrically as
\begin{equation}
R=\frac{A_{\cap}}{A_{\cup}}
\end{equation}
where $A_{\cap}$ denotes the cross section area/volume and $A_{\cup}$
the area/volume of the union of both ellipses/ellipsoids. In order
to align better with the previous defined measures we define the equivalent
complement $\bar{R}=1-R$ instead which becomes zero with a perfect
match.

The goal of this paper is to make a detailed comparison between the
main difference measures proposed so far ($V_{AB}$, $\bar{\gamma}$,
$\mathrm{STRESS}$, $CV$, $\bar{r}$ , $\bar{R}$, $d_{ev}$). We
will in particular compare those difference measures for small deviations
from circularity and we will show that most of them are in fact equivalent.
However, since observed differences are certainly not always infinitesimal
it is also important to study their behavior for large deviations
from circularity and this is where the different proposals thus differ
the most. We will show that $\bar{\gamma}$ is the most sensitive
and $\mathrm{STRESS}$ the least sensitive. We will mostly limit ourselves
to a 2-dimensional chromaticity analysis, but extensions to 3 dimensions
seem straightforward, but calculations may become tedious. In §~(\ref{sec:A-continuous-extension})
we will first explain the continuous extension of the standard discrete
difference measures, eliminating the stochastic effect of the directional
sampling, then in §~\ref{sec:Transformation-based-measures} similar
but simpler transformation based measures will be considered and in
§~(\ref{sec:Eigenvalue-based-difference}) we consider the eigenvalue
based measure.

\section{A continuous extension of the discrete direct color difference measures.}\label{sec:A-continuous-extension}

\subsection{Definitions.}

\begin{figure}
\begin{centering}
\includegraphics[width=10cm]{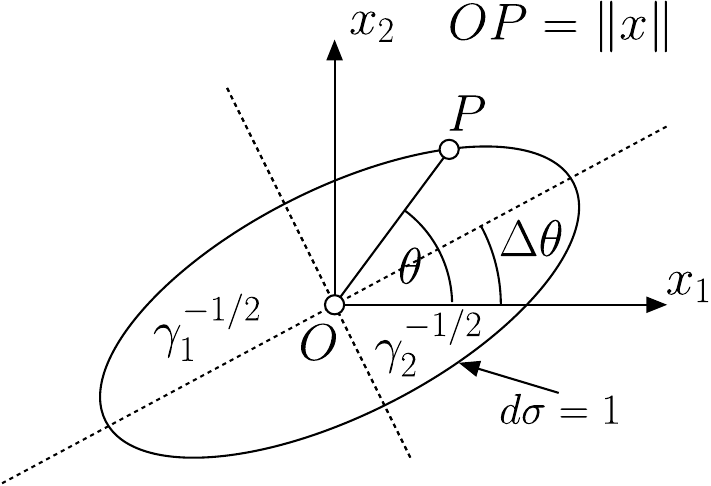}
\par\end{centering}
\caption{A generic (experimental) chromaticity ellipse at threshold in an
arbitrary coordinate system $\left(x_{1},x_{2}\right)$. O is an arbitrary
point of the color manifold and $P$ defines an arbitrary direction
OP, defined by $\theta$. The orientation of the ellipse is defined
by the angle $\Delta\theta$ of the long axis with the $x_{1}$-axis.
The size of the ellipse is defined by the eigenvalues $\gamma_{1}<\gamma_{2}$.}\label{fig:A-generic-chromaticity}
\end{figure}

It is custom to represent an ellipse defined by (\ref{eq:generic LE})
for a perceived difference $d\tau=1$, becoming then a threshold ellipse,
but for any other value of $d\tau$ the ellipse is merely a scaled
version of this threshold ellipse. A generic (chromaticity) ellipse
is shown in Fig.~\ref{fig:A-generic-chromaticity}, using arbitrary
orthonormal coordinates $\left(x^{1},x^{2}\right)$. We will define
the orientation of an ellipse by the angle $\Delta\theta$ made by
its long axis with the $x^{1}$-axis, whereas the long and the short
semi-axes are given by $\gamma^{-1/2}_{i}$, with $\gamma_{1}\leq\gamma_{2}$
the eigenvalues of the metric tensor. The equation of a family of
ellipses can be written using a matrix notation as
\begin{equation}
d\sigma^{2}=x^{\prime}\cdot g\cdot x\label{eq:LE}
\end{equation}
 
where $x$ is the vector $OP$ and the prime indicates the transpose.
Introducing the unit vector along $x$, denoted $\hat{x}$ and with
coordinates $\left(\cos\theta,\sin\theta\right)$, $x=\left\Vert x\right\Vert \hat{x}$,
where $\left\Vert x\right\Vert $ is the Euclidean distance and this
distance can then be written as
\begin{equation}
\left\Vert x\right\Vert =\frac{d\sigma}{\sqrt{\hat{x}^{\prime}\cdot g\cdot\hat{x}}}
\end{equation}
showing that $\left\Vert x\right\Vert $ obviously scales with $d\sigma$.
We note that the variables introduced in the ``Introduction'' $\Delta V_{k}$
and $\Delta E_{k}$ are related by
\begin{equation}
\left\Vert x_{k}\right\Vert =\frac{\Delta V_{k}}{\sqrt{\hat{x}^{\prime}_{k}\cdot g\cdot\hat{x}_{k}}}=\frac{\Delta E_{k}}{\sqrt{\hat{x}^{\prime}_{k}\cdot g^{\mathrm{ref}}\cdot\hat{x}_{k}}}
\end{equation}

since they belong to the same point in color space. We have denoted
the ``experimental'' ellipse belonging to the visual difference
$\Delta V_{k}$ as $g$ and the ``theoretical'' ellipse belonging
to the modeled difference $\Delta E_{k}$, as $g^{\mathrm{ref}}$.
In any standard measure the $\Delta V_{k}$ and $\Delta E_{k}$ values
are thus related by 
\begin{equation}
\frac{\Delta E_{k}}{\Delta V_{k}}=\sqrt{\frac{\hat{x}^{\prime}_{k}\cdot g^{\mathrm{ref}}\cdot\hat{x}_{k}}{\hat{x}^{\prime}_{k}\cdot g\cdot\hat{x}_{k}}}
\end{equation}

However when comparing ellipsoids/ellipses one assumes $d\sigma=\Delta V=\Delta E=1$
and in that case we get
\begin{equation}
\frac{\left\Vert x_{k}\right\Vert _{\Delta V=1}}{\left\Vert x_{k}\right\Vert _{\Delta E=1}}=\sqrt{\frac{\hat{x}^{\prime}_{k}\cdot g^{\mathrm{ref}}\cdot\hat{x}_{k}}{\hat{x}^{\prime}_{k}\cdot g\cdot\hat{x}_{k}}}
\end{equation}

exactly the same, but note the switch in the lhs of these equations.
This suggests that to compare ellipses we should make the substitution
$\Delta E_{k}\rightarrow\left\Vert x_{k}\right\Vert _{V}$ and $\Delta V_{k}\rightarrow\left\Vert x_{k}\right\Vert _{E}$
where we have simplified the notation. Obviously for symmetric measures
like (\ref{eq:VAB}) and (\ref{eq:gamma}) which only depend on the
ratios $\frac{\Delta E_{k}}{\Delta V_{k}}$ and $\frac{\Delta V_{k}}{\Delta E_{k}}$
these substitutions make no difference
\begin{equation}
V^{2}_{AB}=2\left(\sqrt{\left\langle \frac{\left\Vert x\right\Vert _{V}}{\left\Vert x\right\Vert _{E}}\right\rangle \left\langle \frac{\left\Vert x\right\Vert _{E}}{\left\Vert x\right\Vert _{V}}\right\rangle }-1\right)\label{eq:VABcont}
\end{equation}
\begin{equation}
\left(\ln\gamma\right)^{2}=\left\langle \left(\ln\frac{\left\Vert x\right\Vert _{V}}{\left\Vert x\right\Vert _{E}}\right)^{2}\right\rangle -\left\langle \ln\frac{\left\Vert x\right\Vert _{V}}{\left\Vert x\right\Vert _{E}}\right\rangle ^{2}\label{eq:gammacont}
\end{equation}

where the averages are taken over a unit circle and the integrals
involved are considered in §~\ref{subsec:General-continuous-expressions.}.
We accept that they make also sense for the other measures. E.g. the
continuous and symmetric STRESS measure is defined by 
\begin{equation}
\mathrm{STRESS}^{2}=1-\frac{\left\langle \left\Vert x\right\Vert _{E}\left\Vert x\right\Vert _{V}\right\rangle ^{2}}{\left\langle \left\Vert x\right\Vert ^{2}_{V}\right\rangle \left\langle \left\Vert x\right\Vert ^{2}_{E}\right\rangle }\label{eq:STRESScont}
\end{equation}

and for the non-symmetrical $CV$ we find
\begin{equation}
{CV}^{2}=\frac{\left\langle \left\Vert x\right\Vert ^{2}_{V}\right\rangle \left\langle \left\Vert x\right\Vert ^{2}_{E}\right\rangle -\left\langle \left\Vert x\right\Vert _{V}\left\Vert x\right\Vert _{E}\right\rangle ^{2}}{\left\langle \left\Vert x\right\Vert _{V}\right\rangle ^{2}\left\langle \left\Vert x\right\Vert ^{2}_{E}\right\rangle }
\end{equation}
\begin{equation}
{CV}^{2}=\frac{\left\langle \left\Vert x\right\Vert ^{2}_{V}\right\rangle }{\left\langle \left\Vert x\right\Vert _{V}\right\rangle ^{2}}\mathrm{STRESS}^{2}\label{eq:CVcont}
\end{equation}

In a similar way one can derive a continuous version of the $r$ measure
in (\ref{eq:r}) 
\begin{equation}
r=\frac{\left\langle \left\Vert x\right\Vert _{E}\left\Vert x\right\Vert _{V}\right\rangle -\left\langle \left\Vert x\right\Vert _{V}\right\rangle \left\langle \left\Vert x\right\Vert _{E}\right\rangle }{\sqrt{\left(\left\langle \left\Vert x\right\Vert ^{2}_{E}\right\rangle -\left\langle \left\Vert x\right\Vert _{E}\right\rangle ^{2}\right)\left(\left\langle \left\Vert x\right\Vert ^{2}_{V}\right\rangle -\left\langle \left\Vert x\right\Vert _{V}\right\rangle ^{2}\right)}}\label{eq:rcont}
\end{equation}

\subsection{General continuous expressions.}\label{subsec:General-continuous-expressions.}

We will now give the details for evaluating the integrals occurring
in the definitions given in the previous section. We define the 2
ellipsoids to be compared by their eigenvalues ($\lambda_{i}$ for
the model ellipsoid and $\gamma_{i}$ for the experimental one) and
by their relative orientation, that is by a single angle $\Delta\theta$
in the chromatic plane and by 2 angles more generally. We order the
eigenvalues from small to large, thus $\lambda_{1}\leq\lambda_{2}$.
The averages of the single variable quantities $\left\Vert x\right\Vert _{X}$
with $X\in\left\{ V,E\right\} $ depend on the eigenvalues only and
the averages of their product and their ratios depend on all these
quantities. We express the difference between the eigenvalues by a
Michelson type eccentricity
\begin{equation}
\delta_{\lambda}=\frac{1-\frac{\lambda_{1}}{\lambda_{2}}}{1+\frac{\lambda_{1}}{\lambda_{2}}}\quad\frac{\lambda_{1}}{\lambda_{2}}=\frac{1-\delta_{\lambda}}{1+\delta_{\lambda}}\quad\frac{\lambda_{1}+\lambda_{2}}{2}=\frac{\lambda_{1}}{1-\delta_{\lambda}}=\frac{\lambda_{2}}{1+\delta_{\lambda}}
\end{equation}

E.g. for $\delta_{\lambda}=0.5$ the $b/a$ ratio of the ellipse equals
0.58 and drops to 0.33 and 0.23 for $\delta_{\lambda}=0.8$ and $\delta_{\lambda}=0.9$.

We also define a parameter 
\begin{equation}
q_{\lambda}=\frac{\mathrm{tr_{\lambda}}}{2\sqrt{\det_{\lambda}}}=\frac{1}{\sqrt{1-\delta^{2}_{\lambda}}}
\end{equation}
$\mathrm{tr_{\lambda}}=\lambda_{1}+\lambda_{2}$ being the trace of
the tensor and $\det_{\lambda}=\lambda_{1}\lambda_{2}$ its determinant.
The calculation of the averaging integrals occurring in (\ref{eq:CVcont})
and (\ref{eq:rcont}) is straightforward. With the definitions we
get
\begin{equation}
\left\langle \left\Vert x\right\Vert _{E}\right\rangle =\sqrt{\frac{2}{\lambda_{1}+\lambda_{2}}}\frac{2}{\pi}\int^{\frac{\pi}{2}}_{0}\frac{d\theta}{\sqrt{1-\delta_{\lambda}\cos2\theta}}=\frac{2}{\pi}\int^{+\infty}_{0}\frac{dt}{\sqrt{\lambda_{1}t^{2}+\lambda_{2}}\sqrt{1+t^{2}}}
\end{equation}

where the 2nd integral follows with the substitution $\tan\theta=t$.
The first expression is useful in general for numerical integration
but in this case the 2nd expression is readily solved with the result
\begin{equation}
\left\langle \left\Vert x\right\Vert _{E}\right\rangle =\frac{1}{\sqrt{\lambda_{2}}}\frac{2}{\pi}K\left(1-\frac{\lambda_{1}}{\lambda_{2}}\right)\label{eq:xE}
\end{equation}
where $K\left(m\right)$ is the complete Legendre elliptic integral
of the first kind. See Appendix~\ref{sec:Elliptic-Integrals.} for
the notations used. The average of the square is even easier
\begin{equation}
\left\langle \left\Vert x\right\Vert ^{2}_{E}\right\rangle =\frac{2}{\lambda_{1}+\lambda_{2}}\frac{2}{\pi}\int^{\frac{\pi}{2}}_{0}\frac{d\theta}{1-\delta_{\lambda}\cos2\theta}=\frac{2}{\pi}\int^{+\infty}_{0}\frac{dt}{\lambda_{1}t^{2}+\lambda_{2}}=\frac{1}{\sqrt{\lambda_{1}\lambda_{2}}}\label{eq:x2E}
\end{equation}

The results for $\left\langle \left\Vert x\right\Vert _{V}\right\rangle $
and $\left\langle \left\Vert x\right\Vert ^{2}_{V}\right\rangle $
are similar but with the $\lambda$'s replaced by the $\gamma$'s.

The other integrals occurring in the expressions for the different
measures (\ref{eq:STRESScont})(\ref{eq:CVcont})(\ref{eq:VABcont})
and (\ref{eq:gammacont}) can be handled in the same way, except for
the $\gamma$-measure related averages for which no closed form expressions
could be found and these results have been calculated by numerical
integration. These other integrals are more complicated, because we
have to deal with both ellipses at once and only one of them can in
general be aligned with the coordinate axes. Choosing the model one,
with the eigenvalues $\lambda_{i}$, this means that the experimental
metric tensor can still be handled in the same way, but with eigenvalues
$\gamma_{i}$ and with a local angle $\left(\theta-\Delta\theta\right)$
. The average of the product can be written as (scaling the quantity
to make it dimensionless)
\begin{equation}
\frac{\left\langle \left\Vert x\right\Vert _{E}\left\Vert x\right\Vert _{V}\right\rangle }{\sqrt{\left\langle \left\Vert x\right\Vert ^{2}_{E}\right\rangle }\sqrt{\left\langle \left\Vert x\right\Vert ^{2}_{v}\right\rangle }}=\left(1-\delta^{2}_{\lambda}\right)^{1/4}\left(1-\delta^{2}_{\gamma}\right)^{1/4}\frac{1}{\pi}\int^{\frac{\pi}{2}}_{-\frac{\pi}{2}}\frac{d\theta}{\sqrt{1-\delta_{\lambda}\cos2\theta}\sqrt{1-\delta_{\gamma}\cos2\left(\theta-\Delta\theta\right)}}\label{eq:E*V-1}
\end{equation}

and after the substitution $\tan\theta=t$ we obtain 
\begin{equation}
\frac{\left\langle \left\Vert x\right\Vert _{E}\left\Vert x\right\Vert _{V}\right\rangle }{\sqrt{\left\langle \left\Vert x\right\Vert ^{2}_{E}\right\rangle }\sqrt{\left\langle \left\Vert x\right\Vert ^{2}_{v}\right\rangle }}=\frac{\left(1-\delta^{2}_{\lambda}\right)^{1/4}\left(1-\delta^{2}_{\gamma}\right)^{1/4}}{\sqrt{1+\delta_{\lambda}}\sqrt{1+\delta_{\gamma}\cos2\Delta\theta}}I_{0}\left(\delta_{\lambda},\delta_{\gamma},\Delta\theta\right)\label{eq:E*V}
\end{equation}

where
\begin{equation}
I_{0}\left(\delta_{\lambda},\delta_{\gamma},\Delta\theta\right)=\frac{1}{\pi}\int^{+\infty}_{-\infty}\frac{dt}{\sqrt{\left(t^{2}+c^{2}\right)\left(t^{2}+a^{2}-2bt\right)}}\label{eq:I0def}
\end{equation}

where $c$ depends on $\delta_{\lambda}$ and $a,b$ on $\delta_{\gamma}$
and $\Delta\theta.$ This integral is an elliptic one and can be found
using standard procedures which, together with these parameter relations,
are detailed in the Appendix~\ref{sec:Detailed-calculation-of}.
The result, using Eqs. (\ref{eq:E*V}), (\ref{eq:I0result}) and (\ref{eq:nu1result})
can be written succinctly as 
\begin{equation}
\frac{\left\langle \left\Vert x\right\Vert _{E}\left\Vert x\right\Vert _{V}\right\rangle }{\sqrt{\left\langle \left\Vert x\right\Vert ^{2}_{E}\right\rangle }\sqrt{\left\langle \left\Vert x\right\Vert ^{2}_{V}\right\rangle }}=\left(\frac{\nu_{2}}{\nu_{1}}\right)^{1/4}\frac{2}{\pi}R_{F}\left(0,\frac{\nu_{2}}{\nu_{1}},1\right)\label{eq:EVproductresult}
\end{equation}

where $\frac{\nu_{2}}{\nu_{1}}$ is given by (\ref{eq:nu2onu1result}).

The averages of both (dimensionless) ratios can be handled very similarly,
e.g.
\begin{equation}
\left\langle \frac{\left\Vert x\right\Vert _{V}}{\left\Vert x\right\Vert _{E}}\right\rangle =\frac{\left(\lambda_{1}\lambda_{2}\right)^{1/4}}{\left(\gamma_{1}\gamma_{2}\right)^{1/4}}\sqrt{\frac{q_{\lambda}}{q_{\gamma}}}\frac{1}{\pi}\int^{\frac{\pi}{2}}_{-\frac{\pi}{2}}\frac{\sqrt{1-\delta_{\lambda}\cos2\theta}}{\sqrt{1-\delta_{\gamma}\cos2\left(\theta-\Delta\theta\right)}}d\theta\label{eq:VoE1}
\end{equation}
\begin{equation}
\left\langle \frac{\left\Vert x\right\Vert _{V}}{\left\Vert x\right\Vert _{E}}\right\rangle =\frac{\left(\lambda_{1}\lambda_{2}\right)^{1/4}}{\left(\gamma_{1}\gamma_{2}\right)^{1/4}}\sqrt{\frac{q_{\lambda}}{q_{\gamma}}}\sqrt{\frac{1+\delta_{\lambda}}{1+\delta_{\gamma}\cos2\Delta\theta}}I_{1}\left(\delta_{\lambda},\delta_{\gamma},\Delta\theta\right)\label{eq:VoE2}
\end{equation}

where
\begin{equation}
I_{1}\left(\delta_{\lambda},\delta_{\gamma},\Delta\theta\right)=\frac{1}{\pi}\int^{+\infty}_{-\infty}\sqrt{\frac{t^{2}+c^{2}}{t^{2}+a^{2}-2bt}}\frac{dt}{1+t^{2}}\label{eq:I1def}
\end{equation}

Similarly the inverse ratio is given by
\begin{equation}
\left\langle \frac{\left\Vert x\right\Vert _{E}}{\left\Vert x\right\Vert _{V}}\right\rangle =\frac{\left(\gamma_{1}\gamma_{2}\right)^{1/4}}{\left(\lambda_{1}\lambda_{2}\right)^{1/4}}\sqrt{\frac{q_{\gamma}}{q_{\lambda}}}\sqrt{\frac{1+\delta_{\gamma}}{1+\delta_{\lambda}\cos2\Delta\theta}}I^{\prime}_{1}\left(\delta_{\lambda},\delta_{\gamma},\Delta\theta\right)\label{eq:EoV1a}
\end{equation}

where
\begin{equation}
I^{\prime}_{1}\left(\delta_{\lambda},\delta_{\gamma},\Delta\theta\right)=\frac{1}{\pi}\int^{+\infty}_{-\infty}\sqrt{\frac{t^{2}+a^{2}-2bt}{t^{2}+c^{2}}}\frac{dt}{1+t^{2}}\label{eq:Ia1def}
\end{equation}

is similar to integral $I_{1}$ but with numerator and denominator
switched. It can be found most easily by the symmetry relation $I^{\prime}_{1}\left(\delta_{\lambda},\delta_{\gamma},\Delta\theta\right)=I_{1}\left(\delta_{\gamma},\delta_{\lambda},-\Delta\theta\right)=I_{1}\left(\delta_{\gamma},\delta_{\lambda},\Delta\theta\right)$
(Instead of aligning the reference (E) ellipse with the coordinate
axes, we can as well align the experimental (V) ellipse and switch
the sign of $\Delta\theta$, but the latter has no effect.)

Integral $I_{1}$ can be reduced to a combination of integral $I_{0}$
defined in eq.(\ref{eq:I0def}) and a more complicated integral $I_{2}$
\begin{align}
I_{1} & =I_{0}-\left(1-c^{2}\right)I_{2}
\end{align}
 defined by
\begin{equation}
I_{2}\left(\delta_{\lambda},\delta_{\gamma},\Delta\theta\right)=\frac{1}{\pi}\int^{+\infty}_{-\infty}\frac{1}{\sqrt{\left(t^{2}+c^{2}\right)\left(t^{2}+a^{2}-2bt\right)}}\frac{dt}{1+t^{2}}\label{eq:I2def}
\end{equation}

This integral can also be found in closed form as a combination of
symmetric elliptic integrals but one of them with a complex argument
(the details are given in Appendix~(\ref{sec:Detailed-calculation-of})).
With (\ref{eq:I2result}) we obtain

\begin{multline}
I_{1}\left(\delta_{\lambda},\delta_{\gamma},\Delta\theta\right)=\frac{2}{\pi}\frac{1}{\sqrt{\nu_{1}\left(a^{2}-b^{2}\right)}}\left\{ \frac{c^{2}+\nu^{\prime2}_{2}b^{2}}{1+\nu^{\prime2}_{2}b^{2}}R_{F}\left(0,\frac{\nu_{2}}{\nu_{1}},1\right)\right.\\
+\frac{1}{3}\left(1-c^{2}\right)b\left(\nu^{\prime}_{1}+\nu^{\prime}_{2}\right)\frac{\nu^{\prime}_{2}}{\nu^{\prime}_{1}}\left.\Im\left[\frac{b\nu^{\prime}_{1}-i}{\left(b\nu^{\prime}_{2}+i\right)^{3}}R_{J}\left(0,\frac{\nu_{2}}{\nu_{1}},1,-\frac{\nu^{\prime}_{2}}{\nu^{\prime}_{1}}\left(\frac{b\nu^{\prime}_{1}-i}{b\nu^{\prime}_{2}+i}\right)^{2}\right)\right]\right\} \label{eq:I1result}
\end{multline}

\begin{figure}
\begin{centering}
\includegraphics[width=10cm]{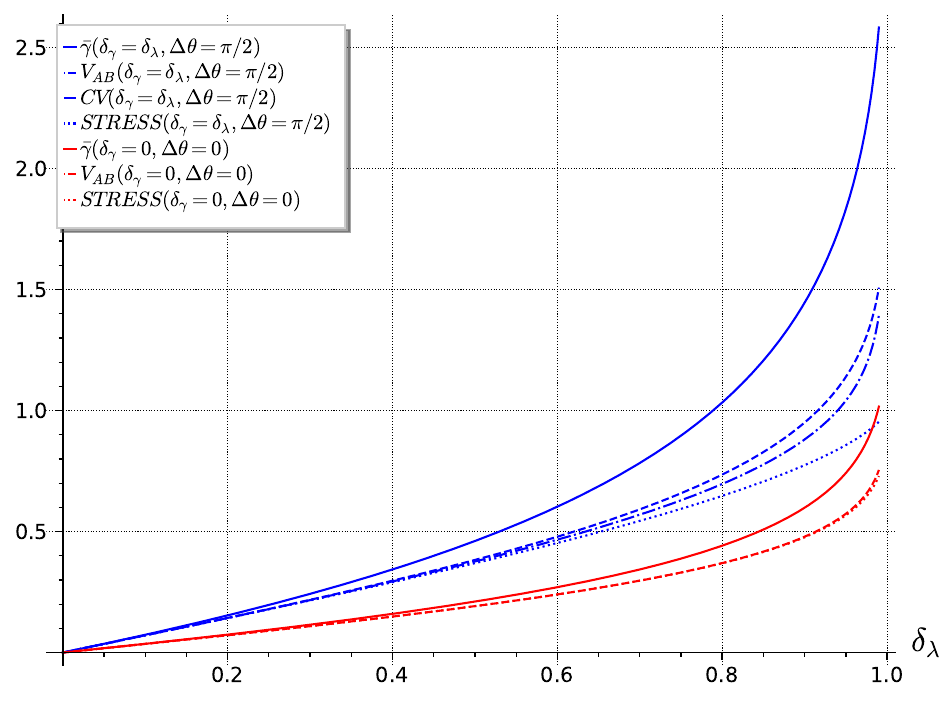}
\par\end{centering}
\caption{The continuous difference measures $\bar{\gamma}$ (solid), $V_{AB}$
(dash) , $\mathrm{STRESS}$ (dot) and $CV$ (dashdot) as a function
of $\delta_{\lambda}$ for (i) $\delta_{\gamma}=\delta_{\lambda}$
and $\Delta\theta=\pi/2$ (blue curves, this is 2 equal ellipses but
perpendicular to each other) and for (ii) $\delta_{\gamma}=0$ and
$\Delta\theta=0$ (red curves, this compares an ellipse with a circle).
In the 2nd case $\mathrm{STRESS}$ and $CV$ are identical and $V_{AB}$
almost identical. For the first case the 3 measures $V_{AB}$, $\mathrm{STRESS}$
and $CV$ are also very similar at least up to $\delta=0.9$, corresponding
with an aspect ratio of $1/\sqrt{19}=0.23$.}\label{fig:The-difference-measures}
\end{figure}

The results of these difference measures are illustrated in Fig.\ref{fig:The-difference-measures}.
The difference surfaces $d\left(\delta_{\lambda},\delta_{\gamma}\right)$
with $\Delta\theta$ as a parameter are symmetric in $\delta_{\lambda},\delta_{\gamma}$.
The difference is zero along the diagonal $\delta_{\lambda}=\delta_{\gamma}$
for $\Delta\theta=0$ but maximal for $\Delta\theta=\pi/2$. For small
deviations the measures are all identical. For large deviations the
$\bar{\gamma}$ measure gives the largest value, followed by $V_{AB}$
and the smallest ones are given by the STRESS measure. The difference
between $V_{AB}$ and $\mathrm{STRESS}$ is smaller than between $\bar{\gamma}$
and $V_{AB}$. For all practical purposes $V_{AB}$ and $CV$ are
very similar and since all measures are equivalent for small deviations
from circularity the real choice is between the most sensitive one
$\left(\bar{\gamma}\right)$ and the least sensitive one $\left(\mathrm{STRESS}\right)$.

\subsection{Discrete difference measures.}

\begin{figure}
\begin{centering}
\includegraphics[width=10cm]{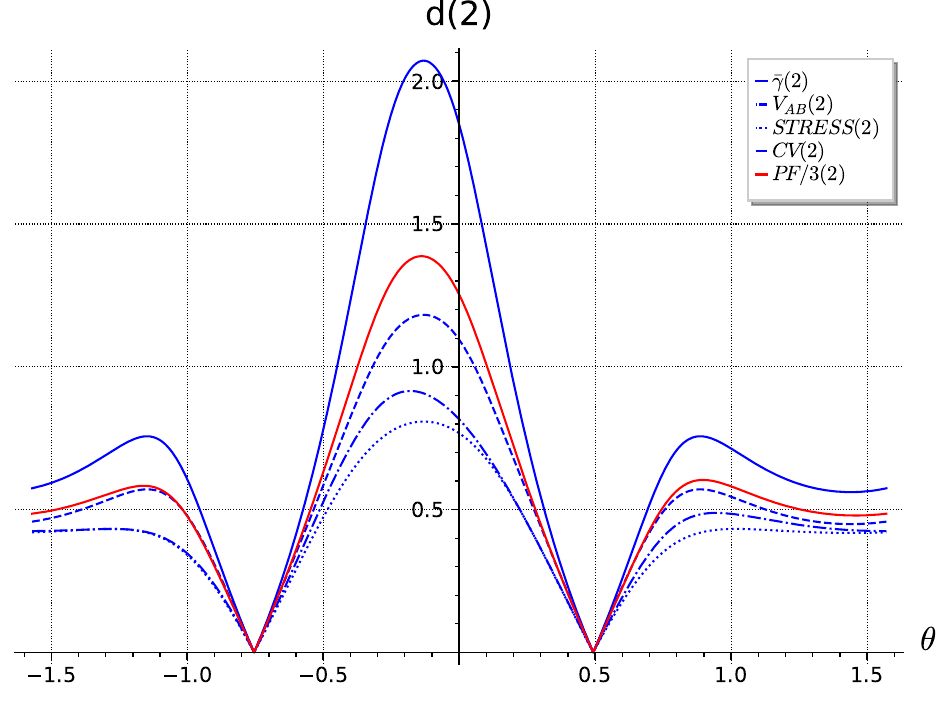}
\par\end{centering}
\caption{Discrete difference measures $\bar{\gamma}$ (solid), $V_{AB}$ (dash)
, $\mathrm{STRESS}$ (dot) and $CV$ (dashdot) for an absolute minimum
of 2 sampled directions $\frac{\pi}{3}$ apart ($\theta$ and $\theta+\pi/3)$
and for 2 rather elongated ellipses ($\delta_{\lambda}=\delta_{\gamma}=0.9$
, $\Delta\theta=\pi/4$). The red curve shows the average measure
$PF/3$.}\label{fig:Discrete-difference-measures}
\end{figure}

There is one situation where these continuous measures might not be
advisable and the original discrete ones can still be applied. That
is when the number of directions in a color point is limited and the
reliable definition of an ellipsoid is impossible or at least unreliable.
In that case it still makes sense to compare those limited data with
the discrete measures. To compare those measures we consider an extreme
case of very elongated ellipses ($\delta_{\lambda}=\delta_{\gamma}=0.9$)
at an angle $\Delta\theta=\pi/4$. We also consider an absolute minimum
of only two sampled directions, arbitrarily $\frac{\pi}{3}$ apart,
since for a single direction the (symmetric) measures always return
a difference 0. As expected (see Fig.\ref{fig:Discrete-difference-measures})
the different measures are very similar. They tend to each other when
the differences are small and they differ most when the differences
become large. We note that notwithstanding the large difference between
these ellipses the discrete error nevertheless vanishes for 2 particular
directions $\theta$ and these nodes persist even if $\delta_{\lambda}\neq\delta_{\gamma}$.
These observations make the use of the averaged measure $PF/3$ rather
moot and we believe that the inadequate sampling of the directions
in a color point is responsible for the observed scattering of the
differences. Finally we notice that also here the $\bar{\gamma}$
measure is the most sensitive and STRESS the least one. Comparing
Fig.\ref{fig:Discrete-difference-measures} with Fig.\ref{fig:The-difference-measures}
we can conclude that for these extreme sampling scenarios the 4 measures
considered are very similar and the only criterium which matters is
what sensitivity is wanted.

\subsection{Expansions of the continuous measures for small deviations from circularity.}

We consider the expansions of these continuous measures in the limit
for $\delta_{\lambda}\rightarrow0$ and $\delta_{\gamma}\rightarrow0$
meaning that the ellipses deviate little from circularity. In that
case the angle $\Delta\theta$ is of no great importance and can be
arbitrary. We could also consider arbitrary ellipses which are almost
equal and then also this angle should remain small. This is the limit
for $b\rightarrow0$. We will consider this case in §~\ref{sec:Transformation-based-measures}.

Using basic series expansions the integrands in (\ref{eq:xE})-(\ref{eq:E*V-1}),(\ref{eq:VoE1}),(\ref{eq:EoV1a})
can be expanded and these integrals are then easily obtained in closed
form. Since linear terms integrate to zero the expansions must include
at least quadratic terms yielding
\begin{equation}
\frac{\left\langle \left\Vert x\right\Vert _{E}\left\Vert x\right\Vert _{V}\right\rangle }{\sqrt{\left\langle \left\Vert x\right\Vert ^{2}_{E}\right\rangle \left\langle \left\Vert x\right\Vert ^{2}_{V}\right\rangle }}\approx1-\frac{1}{16}\left(\delta^{2}_{\lambda}+\delta^{2}_{\gamma}-2\delta_{\lambda}\delta_{\gamma}\cos2\Delta\theta\right)
\end{equation}
\begin{equation}
\left\langle \frac{\left\Vert x\right\Vert _{V}}{\left\Vert x\right\Vert _{E}}\right\rangle \approx1+\frac{1}{16}\left(3\delta^{2}_{\lambda}-\delta^{2}_{\gamma}-2\delta_{\lambda}\delta_{\gamma}\cos2\Delta\theta\right)
\end{equation}
\begin{equation}
\left\langle \frac{\left\Vert x\right\Vert _{E}}{\left\Vert x\right\Vert _{V}}\right\rangle \approx1+\frac{1}{16}\left(3\delta^{2}_{\gamma}-\delta^{2}_{\lambda}-2\delta_{\lambda}\delta_{\gamma}\cos2\Delta\theta\right)
\end{equation}
\begin{equation}
\left\langle \ln\frac{\left\Vert x\right\Vert _{V}}{\left\Vert x\right\Vert _{E}}\right\rangle \approx\frac{1}{8}\left(\delta^{2}_{\lambda}-\delta^{2}_{\gamma}\right)
\end{equation}
\begin{equation}
\left\langle \left(\ln\frac{\left\Vert x\right\Vert _{V}}{\left\Vert x\right\Vert _{E}}\right)^{2}\right\rangle \approx\frac{1}{8}\left(\delta^{2}_{\gamma}+\delta^{2}_{\lambda}-2\delta_{\gamma}\delta_{\lambda}\cos2\Delta\theta\right)
\end{equation}

With these we then find the series expansions for the difference metrics
(\ref{eq:VABcont})-(\ref{eq:CVcont})
\begin{equation}
V_{AB}\approx\frac{1}{2\sqrt{2}}\sqrt{\delta^{2}_{\lambda}+\delta^{2}_{\gamma}-2\delta_{\lambda}\delta_{\gamma}\cos2\Delta\theta}\label{eq:VABsmalldelta}
\end{equation}
\begin{equation}
\ln\gamma\approx\bar{\gamma}\approx\frac{1}{2\sqrt{2}}\sqrt{\delta^{2}_{\gamma}+\delta^{2}_{\lambda}-2\delta_{\gamma}\delta_{\lambda}\cos2\Delta\theta}\label{eq:lngsmalldelta}
\end{equation}
\begin{equation}
\mathrm{STRESS}\approx\frac{1}{2\sqrt{2}}\sqrt{\delta^{2}_{\lambda}+\delta^{2}_{\gamma}-2\delta_{\lambda}\delta_{\gamma}\cos2\Delta\theta}\label{eq:stresssmalldelta}
\end{equation}
\begin{equation}
CV\approx\frac{\sqrt{\left\langle \left\Vert x\right\Vert ^{2}_{V}\right\rangle }}{\left\langle \left\Vert x\right\Vert _{V}\right\rangle }\mathrm{STRESS}\label{eq:CVsmalldelta1}
\end{equation}

where
\begin{equation}
\frac{\sqrt{\left\langle \left\Vert x\right\Vert ^{2}_{V}\right\rangle }}{\left\langle \left\Vert x\right\Vert _{V}\right\rangle }\approx1-\frac{5}{2}\delta_{\gamma}+\frac{3}{16}\delta^{2}_{\gamma}\approx1\label{eq:CDsmalldelta2}
\end{equation}

Apparently these four difference metrics are all equivalent in the
limit for small deviations from circularity and up to the 2nd order
in the $\delta$'s.

\subsection{The correlation coefficient.}\label{subsec:The-correlation-coefficient.}

\begin{figure}
\begin{centering}
\includegraphics[width=10cm]{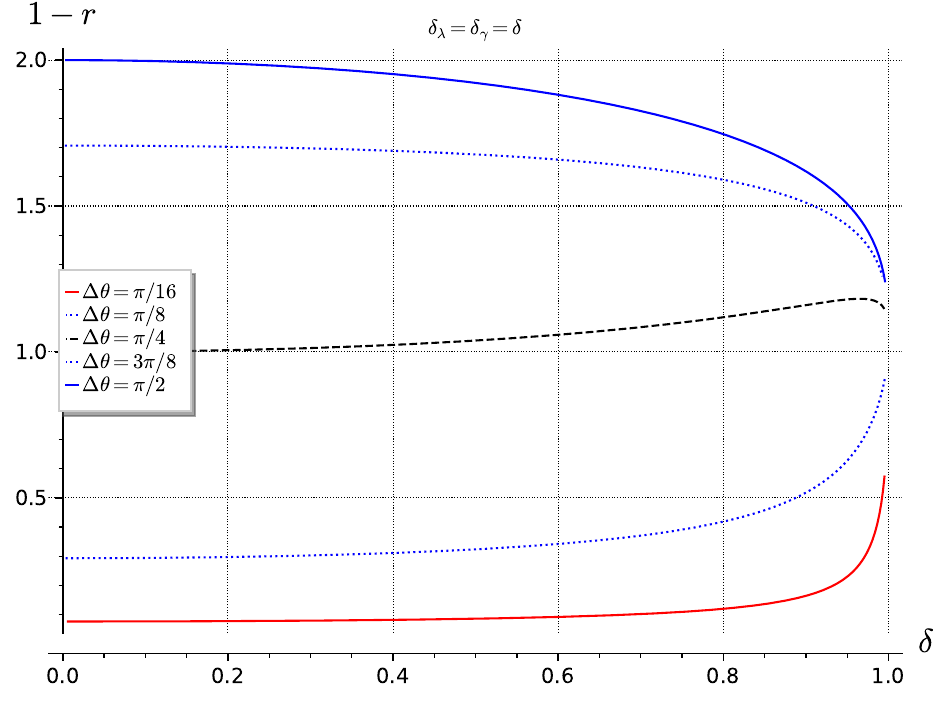}
\par\end{centering}
\caption{The continuous correlation coefficient $\bar{r}=1-r$ as a function
of $\delta_{\lambda}=\delta_{\gamma}$ for $\Delta\theta=\frac{\pi}{16},\frac{\pi}{8},\frac{\pi}{4},\frac{3\pi}{8}$
and $\frac{\pi}{2}$. The blue solid curve is directly comparable
with the blue curves in Fig.\ref{fig:The-difference-measures}.}\label{fig:The-continuous-correlation}
\end{figure}

If we apply the same continuous extension to the (Pearson) correlation
coefficient defined in (\ref{eq:r}) we get the result (\ref{eq:rcont})
which is shown in Fig.\ref{fig:The-continuous-correlation}. Although
$r$ is superficially related to the $\mathrm{STRESS}$ measure it
shows, primarily due to the denominator, a complete different behavior.
Contrary to all previous considered measures, $r$ does not tend to
zero when $\delta\rightarrow0$. In that limit the ellipses are almost
circular and for all practical purposes their orientation should become
quite irrelevant. But in this limit $r$ is instead supersensitive
for the orientation difference $\Delta\theta$ since in fact
\begin{equation}
\lim_{\delta_{\lambda},\delta_{\gamma}\rightarrow0}r=\cos2\Delta\theta\left[1-\frac{23}{64}\left(\delta^{2}_{\lambda}+\delta^{2}_{\gamma}\right)\right]+\frac{9}{64}\delta_{\lambda}\delta_{\gamma}\cos4\Delta\theta+O\left(\delta^{4}\right)\label{eq:rlimitsmalldelta}
\end{equation}

and $r\left(0,0,\Delta\theta\right)=\cos2\Delta\theta.$ Accompanying
this orientation sensitivity we notice the relative insensitivity
to $\delta$ and thus the aspect ratio, the curves staying relatively
flat unless $\delta$ approaches unity. When the ellipses become very
elongated ($\delta\rightarrow1$) the orientation has not longer any
influence at all, $r$ becoming zero here (but $\bar{r}=1$ halfway
between its maximum and zero). In a sense $\bar{r}=1-r$ on the one
hand and the $\bar{\gamma},V_{AB},CV$ and STRESS measures on the
other hand are complementary with the first one being supersensitive
to orientation mismatches and not to $\delta$, while the others are
linear in $\delta$ and rather insensitive to orientation mismatches.
Very likely Luo's remark referred to in the ``Introduction'' about
`` (r)... being inconsistent with the other measures ...'' refers
to this markedly different behavior. In their paper Kirchner e.a.
\cite{Kirchner:2011zy} emphasize that this different behavior is
due to the fact that STRESS forces the model regression line to pass
through the origin whereas $\bar{r}$ does not. Unfortunately this
has the consequence that $\bar{r}$ becomes coordinate dependent.
When comparing a circle ($\delta_{\lambda}=0)$ with an arbitrary
ellipse ($\delta_{\gamma}\neq0$) the angle $\Delta\theta$ has no
physical meaning and only reflects the choice of the coordinates used.
Nevertheless, as shown by (\ref{eq:rlimitsmalldelta}), $\bar{r}$
still depends on this (arbitrary) angle. The other measures behave
according with (\ref{eq:deltamu2}) where the angle dependence disappears
if one of the $\delta$'s becomes zero. Carter e.a. \cite{Carter:2022nr}
advise the use of centered data and show that in that case $r_{c}=r$
is equivalent with $\mathrm{STRESS}$, which is indeed obvious if
Eqs.\ref{eq:STRESScont} and \ref{eq:rcont} are compared (with the
averages zeros). But STRESS makes sense for not-centered data and
if we should use centered data instead STRESS will inherit the same
unwanted property of $r$.

\section{Transformation based measures.}\label{sec:Transformation-based-measures}

\begin{figure}
\begin{centering}
\includegraphics[width=10cm]{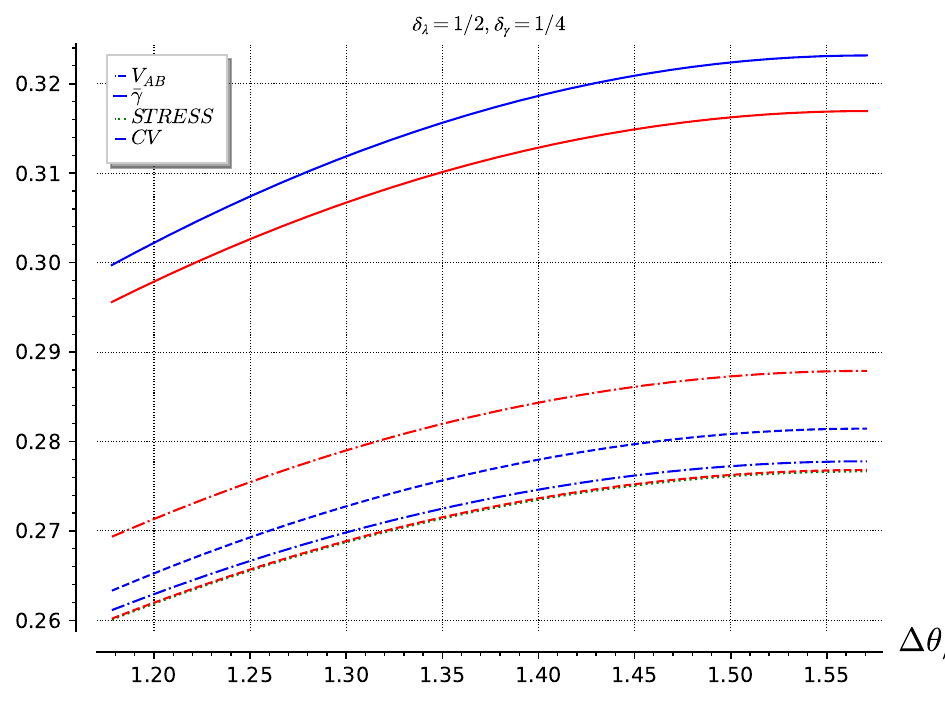}
\par\end{centering}
\caption{Comparison between the general continuous measures (in blue) $\bar{\gamma}$
(solid), $V_{AB}$ (dash), STRESS (dot) and $CV$ (dashdot) and the
transformation based ones (in red) as a function of $\frac{3\pi}{8}<\Delta\theta<\frac{\pi}{2}$
for arbitrary values $\delta_{\lambda}=1/2$ and $\delta_{\gamma}=1/4$.
The unique curve STRESS is almost indistinguishable from the transformation
based $V_{AB,\mu}$ curve.}\label{fig:coordinateIndependence}
\end{figure}

As explained in the Introduction, when comparing ellipsoids the difference
metric should ideally not depend on the particular coordinates chosen,
since it should be an intrinsic property of the tensors to be compared.
The effects of coordinate dependence are illustrated in Fig.\ref{fig:coordinateIndependence}.
It is then advantageous to transform both ellipsoids to coordinates
into which one of the ellipsoids becomes spherical since the comparison
becomes much simpler. Obviously the correlation measure $\bar{r}$
cannot be used for this purpose, since it's exactly for this situation
that $\bar{r}$ is ill-defined. We believe this was the strategy followed
by MacAdam \cite{MacAdam:1964hk}, but as far as we know his method
had little followers. With $g$ the tensor of the experimental ellipsoid
and $g^{\mathrm{ref}}$ that of the theoretical one, we then consider
the transformed ellipsoids defined by $h^{\mathrm{ref}}=I$ and $h=\left(g^{\mathrm{ref}}\right)^{-\frac{1}{2}}\cdot g\cdot\left(g^{\mathrm{ref}}\right)^{-\frac{1}{2}}$,
where the splitting of the inverse of the reference tensor into two
equal square roots is required to guarantee that $h$ remains symmetric
and this splitting does not affect the eigenvalues of $h$, being
the same as those of $g\cdot\left(g^{\mathrm{ref}}\right)^{-1}$.
If both ellipsoids are identical, $h$ reduces also to the unit tensor.
Given the eigenvalues of $g$ and $g^{\mathrm{ref}}$ and their relative
orientation ($\delta_{\lambda}$, $\delta_{\gamma}$, $\Delta\theta$)
the eigenvalues of $h$ can be obtained with ($\det h=\mu_{1}\mu_{2}=\frac{\gamma_{1}\gamma_{2}}{\lambda_{1}\lambda_{2}}$,
$\mathrm{tr}h=\mu_{1}+\mu_{2}$)
\begin{equation}
q_{\mu}=\frac{\mathrm{tr}h}{2\sqrt{\det h}}=q_{\lambda}q_{\gamma}\left(1-\delta_{\lambda}\delta_{\gamma}\cos2\Delta\theta\right)\label{eq:deltaq}
\end{equation}
\begin{equation}
\delta^{2}_{\mu}=1-\frac{1}{q^{2}_{\mu}}\label{eq:deltamu}
\end{equation}

Whereas MacAdam sampled the resulting ellipsoid, we can apply the
continuous method presented in §~\ref{sec:A-continuous-extension}.
Because the reference ellipsoid is now spherical we set $\lambda_{i}=1$
($\left\Vert x\right\Vert _{E}=1$, $c=1$) and in addition the angle
$\Delta\theta$ is inconsequential and can also be set zero ($\Delta\theta=0$,
$b=0$). With $\mu_{i}$ the eigenvalues of $h$ ($a^{2}=\frac{\mu_{1}}{\mu_{2}}\leq1$)
we obtain (To avoid confusion with the continuous measures already
defined, we add an index $\mu$ to label these transformation based
but still continuous measures. However we use no specific notation
for $a,b,c$.)
\begin{equation}
\left\langle \left\Vert x\right\Vert _{V,\mu}\right\rangle =\frac{1}{\sqrt{\mu_{2}}}\frac{2}{\pi}R_{F}\left(0,\frac{\mu_{1}}{\mu_{2}},1\right)=\frac{1}{\sqrt{\mu_{2}}}\frac{2}{\pi}K\left(1-\frac{\mu_{1}}{\mu_{2}}\right)
\end{equation}
\begin{equation}
\left\langle \frac{1}{\left\Vert x\right\Vert _{V,\mu}}\right\rangle =\sqrt{\mu_{2}}\frac{2}{\pi}2R_{G}\left(0,\frac{\mu_{1}}{\mu_{2}},1\right)=\sqrt{\mu_{2}}\frac{2}{\pi}E\left(1-\frac{\mu_{1}}{\mu_{2}}\right)
\end{equation}
 {}
\begin{equation}
\left\langle \left\Vert x\right\Vert ^{2}_{V,\mu}\right\rangle =\frac{1}{\sqrt{\mu_{1}\mu_{2}}}
\end{equation}
 {}
\begin{align}
V^{2}_{AB,\mu} & =2\left(\frac{2}{\pi}\sqrt{R_{F}\left(0,\frac{\mu_{1}}{\mu_{2}},1\right)2R_{G}\left(0,\frac{\mu_{1}}{\mu_{2}},1\right)}-1\right)\label{eq:VABmu}
\end{align}
\begin{align}
\mathrm{STRESS^{2}_{\mu}} & =1-\sqrt{\frac{\mu_{1}}{\mu_{2}}}\left[\frac{2}{\pi}R_{F}\left(0,\frac{\mu_{1}}{\mu_{2}},1\right)\right]^{2}\label{eq:STRESS2mu}
\end{align}
\begin{align}
{CV}_{\mu} & =\frac{\mathrm{STRESS}}{\sqrt{1-\mathrm{STRESS}^{2}}}\label{eq:CVmu}
\end{align}

The $\gamma$ related quantities still can not be found completely
in closed form
\begin{equation}
-\left\langle \ln\left\Vert x\right\Vert ^{2}_{V,\mu}\right\rangle =\ln\frac{\mu_{1}+\mu_{2}}{2}+\frac{2}{\pi}\int^{\pi/2}_{0}\ln\left(1-\delta_{\mu}\cos2\theta\right)d\theta
\end{equation}
\begin{multline}
\left\langle \left(\ln\left\Vert x\right\Vert ^{2}_{V,\mu}\right)^{2}\right\rangle =\left(\ln\frac{\mu_{1}+\mu_{2}}{2}\right)^{2}+2\ln\frac{\mu_{1}+\mu_{2}}{2}\frac{2}{\pi}\int^{\pi/2}_{0}\ln\left(1-\delta_{\mu}\cos2\theta\right)d\theta\\
+\frac{2}{\pi}\int^{\pi/2}_{0}\left[\ln\left(1-\delta_{\mu}\cos2\theta\right)\right]^{2}d\theta
\end{multline}
\begin{equation}
\left(\ln\gamma_{\mu}\right)^{2}=\frac{1}{4}\left[\left\langle \left(\ln\left\Vert x\right\Vert ^{2}_{V,\mu}\right)^{2}\right\rangle -\left\langle \ln\left\Vert x\right\Vert ^{2}_{V,\mu}\right\rangle ^{2}\right]\label{eq:gamma cont transf}
\end{equation}
\begin{equation}
\left(\ln\gamma_{\mu}\right)^{2}=\frac{1}{2\pi}\int^{\pi/2}_{0}\left[\ln\left(1-\delta_{\mu}\cos2\theta\right)\right]^{2}d\theta-\frac{1}{\pi^{2}}\left[\int^{\pi/2}_{0}\ln\left(1-\delta_{\mu}\cos2\theta\right)d\theta\right]^{2}
\end{equation}

Although the first moment has a closed expression
\begin{equation}
\frac{1}{\pi}\int^{\pi/2}_{0}\ln\left(1-\delta_{\mu}\cos2\theta\right)d\theta=\ln\frac{\sqrt{1-\delta_{\mu}}+\sqrt{1+\delta_{\mu}}}{2}
\end{equation}
we couldn't find one for the 2nd moment.

\begin{figure}
\begin{centering}
\includegraphics[width=10cm]{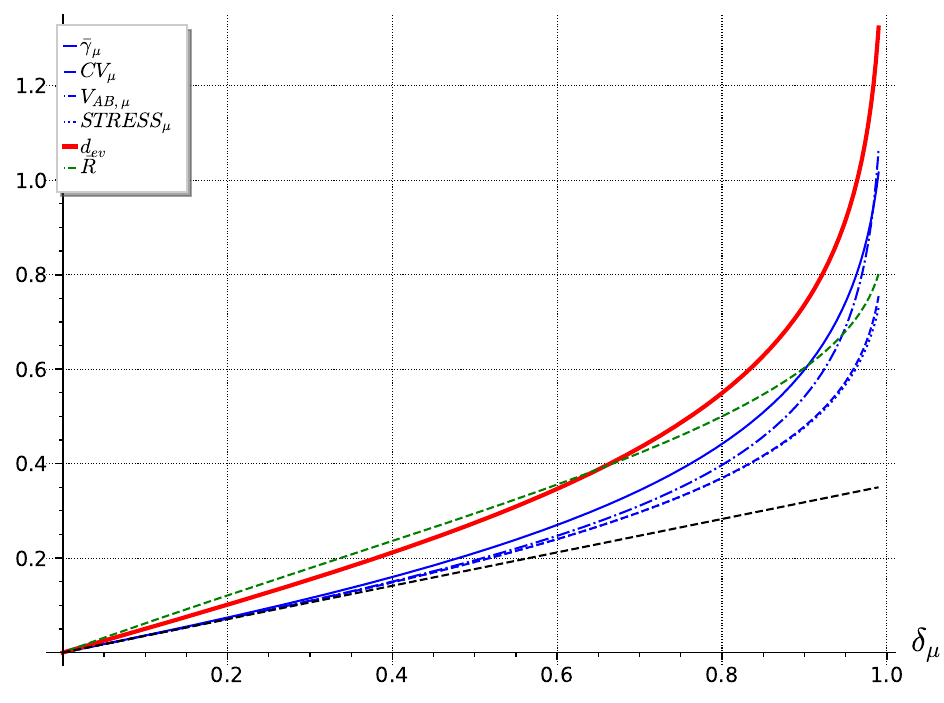}
\par\end{centering}
\caption{The difference measures $\bar{\gamma}_{\mu}$ (solid), $V_{AB,\mu}$
(dash) , $\mathrm{STRESS}$ (dot) ${CV}_{\mu}$ (dashdot) $\bar{R}$
(dashed green line) and $d_{ev}$ (heavy red line) as a function of
$\delta_{\mu}$ for ellipses. $V_{AB,\mu}$ and $\mathrm{STRESS}$
are almost coincident. The straight line is the mutual asymptote of
the first 4 measures ($\frac{\delta_{\mu}}{2\sqrt{2}}$) for $\delta_{\mu}\rightarrow0$.
The slopes of $d_{ev}$ ($\frac{\delta_{\mu}}{2}$) and of $\bar{R}$
($\frac{2}{\pi}\delta_{\mu}$) are larger by factors $\sqrt{2}\approx1.4$
and $\frac{4\sqrt{2}}{\pi}\approx1.8$. Note that $d_{ev}$, $\mathrm{STRESS}$
and $\bar{R}$ are coordinate independent.}\label{fig:The-difference-measures-2}
\end{figure}

All these measures are functions of a single argument $\delta_{\mu}=\frac{\mu_{2}-\mu_{1}}{\mu_{2}+\mu_{1}}$.
Those dependencies of $\delta_{\mu}$ are shown in Fig.\ref{fig:The-difference-measures-2}.
For small $\delta_{\mu}$ these measures are identical with the expansion
\begin{equation}
\lim_{\delta_{\gamma}\rightarrow0}\begin{bmatrix}\bar{\gamma}_{\mu}\\
{CV}_{\mu}\\
V_{AB,\mu}\\
\mathrm{STRESS}_{\mu}
\end{bmatrix}=\frac{\delta_{\mu}}{2\sqrt{2}}\label{eq:small limit}
\end{equation}

but also for greater differences they are very similar. From (\ref{eq:deltaq})
$\delta_{\mu}$ can be calculated in function of $\delta_{\lambda}$,
$\delta_{\gamma}$ and $\Delta\theta$ and in the limit for $\left(\delta_{\lambda},\delta_{\gamma}\right)\rightarrow0$
we find 
\begin{equation}
\lim_{\delta_{\lambda},\delta_{\gamma}\rightarrow0}\delta^{2}_{\mu}=\delta^{2}_{\lambda}+\delta^{2}_{\gamma}-2\delta_{\lambda}\delta_{\gamma}\cos2\Delta\theta\label{eq:deltamu2}
\end{equation}

explaining the dependencies observed in (\ref{eq:VABsmalldelta})(\ref{eq:lngsmalldelta})(\ref{eq:stresssmalldelta})(\ref{eq:CVsmalldelta1})(\ref{eq:CDsmalldelta2}).
In the other extreme ($\delta_{\mu}\rightarrow1$, $\frac{\mu_{1}}{\mu_{2}}\rightarrow0$)
it follows from Eq.(\ref{eq:VABmu})(\ref{eq:STRESS2mu}) that $\mathrm{STRESS}_{\mu}$
and $V_{AB,\mu}$ are also very similar and for all practical purposes
we can set $\mathrm{STRESS}_{\mu}\approx V_{AB,\mu}$.

Concerning the coordinate independence of $\bar{\gamma}$, $V_{AB}$,
STRESS and $CV$ we compare the continuous measures with the transformation
based ones in Fig.~(\ref{fig:coordinateIndependence}) for 2 ellipses
with constant aspect ratios and as a function of the angle between
their long axes. Except for the STRESS measure all the others are
coordinate dependent. The general continuous STRESS result is given
by (\ref{eq:STRESScont}) together with (\ref{eq:EVproductresult})
and the transformation based $\mathrm{STRESS}_{\mu}$ result by (\ref{eq:STRESS2mu}).
They are indeed identical because $\frac{\nu_{2}}{\nu_{1}}=\frac{\mu_{1}}{\mu_{2}}$
where the latter follows from (\ref{eq:deltamu}) and the former was
given in (\ref{eq:nu2onu1result}). The roots $\nu_{i}$ defined for
reducing the elliptic integrals thus coincide with the eigenvalues
of the tensor $h=\left(g^{\mathrm{ref}}\right)^{-\frac{1}{2}}\cdot g\cdot\left(g^{\mathrm{ref}}\right)^{-\frac{1}{2}}$.
This equality also holds for the other measures e.g. in (\ref{eq:I1result})
occurring in $V_{AB}$ but the unique property of STRESS is that it
depends on $\frac{\nu_{2}}{\nu_{1}}$ only whereas the other measures
depend on other quantities too. E.g. $CV_{\mu}$ is merely a variation
of $\mathrm{STRESS}_{\mu}$ as evidenced by (\ref{eq:CVmu}) but although
$CV$ also depends on STRESS (see (\ref{eq:CVcont})) the relation
is different. This lack of coordinate independence is restricted to
relatively large values of the aspect ratio since for small deviations
from circularity ($\delta_{\lambda},\delta_{\gamma}\rightarrow0$)
all measures are coordinate independent as evidenced by (\ref{eq:deltamu}).

\section{Eigenvalue based difference measures.}\label{sec:Eigenvalue-based-difference}

The transformation based measures considered in the previous section
compare the ellipsoid defined by $h=\left(g^{\mathrm{ref}}\right)^{-\frac{1}{2}}\cdot g\cdot\left(g^{\mathrm{ref}}\right)^{-\frac{1}{2}}$
with the unit ball instead of comparing the ellipsoids defined by
$g$ and $g^{\mathrm{ref}}$ directly. Arrived at this point it seems
more efficient to define a measure based on the eigenvalues $\mu_{i}$
of the tensor $h$. The straightforward rms measure $\left\langle \left(\sqrt{\mu_{i}}-1\right)^{2}\right\rangle ^{1/2}$
being asymmetrical, we have chosen \cite{Candry:2022kx} for the definition
\begin{equation}
d^{2}_{ev}=\left\langle \left(\ln\sqrt{\mu_{i}}\right)^{2}\right\rangle \label{eq:d-base}
\end{equation}

which gives the same result $\frac{1}{4}\left\langle \epsilon^{2}_{i}\right\rangle $for
small deviations from the unit ball with $\mu_{i}=1+\epsilon_{i}$.
Since (\ref{eq:d-base}) is not scale invariant we scale the $\mu_{i}\rightarrow F\mu_{i}$
and choose $F$ optimally, that is
\begin{equation}
\ln F_{\mathrm{opt}}=-\left\langle \ln\mu_{i}\right\rangle 
\end{equation}
\begin{equation}
d^{2}_{\mathrm{opt}}=d^{2}_{ev}=\frac{1}{4}\left[\left\langle \left(\ln\mu_{i}\right)^{2}\right\rangle -\left\langle \ln\mu_{i}\right\rangle ^{2}\right]\label{eq:dev}
\end{equation}

Apparently $d_{ev}$ is the eigenvalue counterpart of what $\ln\gamma_{\mu}$
is for the continuous measure as shown in (\ref{eq:gamma cont transf})
but $d_{ev}/\sqrt{2}$ is actually nearer to ${CV}_{\mu}$ than to
$\bar{\gamma}_{\mu}$ (see Fig.(\ref{fig:The-difference-measures-2})).

The expression in (\ref{eq:dev}) holds for 2-dimensional (chromatic)
ellipses as well as for 3-dimensional ellipsoids. In the first case
one readily derives that
\begin{equation}
d_{ev}=\frac{1}{4}\left|\ln\frac{\mu_{1}}{\mu_{2}}\right|=\frac{1}{4}\left|\ln\frac{1-\delta_{\mu}}{1+\delta_{\mu}}\right|
\end{equation}

with the limit
\begin{equation}
\lim_{\delta_{\mu}\rightarrow0}d_{ev}=\frac{\delta_{\mu}}{2}+O\left(\delta^{3}_{\mu}\right)\label{eq:dev2dsmall}
\end{equation}

and this eigenvalue based measured is $\sqrt{2}$ times the previously
defined measures as shown in (\ref{eq:small limit}).

For an ellipsoid $d_{ev}$ can be expressed as a function of two eccentricities
$\frac{\mu_{1}}{\mu_{2}}=\frac{1-\delta_{12}}{1+\delta_{12}}$ and
$\frac{\mu_{2}}{\mu_{3}}=\frac{1-\delta_{23}}{1+\delta_{23}}$ with
$\mu_{1}\leq\mu_{2}\leq\mu_{3}$
\begin{equation}
d^{2}_{ev}=\frac{1}{18}\left[\left(\ln\frac{\mu_{1}}{\mu_{2}}\right)^{2}+\left(\ln\frac{\mu_{2}}{\mu_{3}}\right)^{2}+\ln\frac{\mu_{1}}{\mu_{2}}\ln\frac{\mu_{2}}{\mu_{3}}\right]
\end{equation}

with as limit
\begin{equation}
\lim_{\delta_{12},\delta_{23}\rightarrow0}d_{ev}=\frac{\sqrt{2}}{3}\sqrt{\delta^{2}_{12}+\delta^{2}_{23}+\delta_{12}\delta_{23}}+O\left(\delta^{3}\right)
\end{equation}

\begin{figure}
\begin{centering}
\includegraphics[width=10cm]{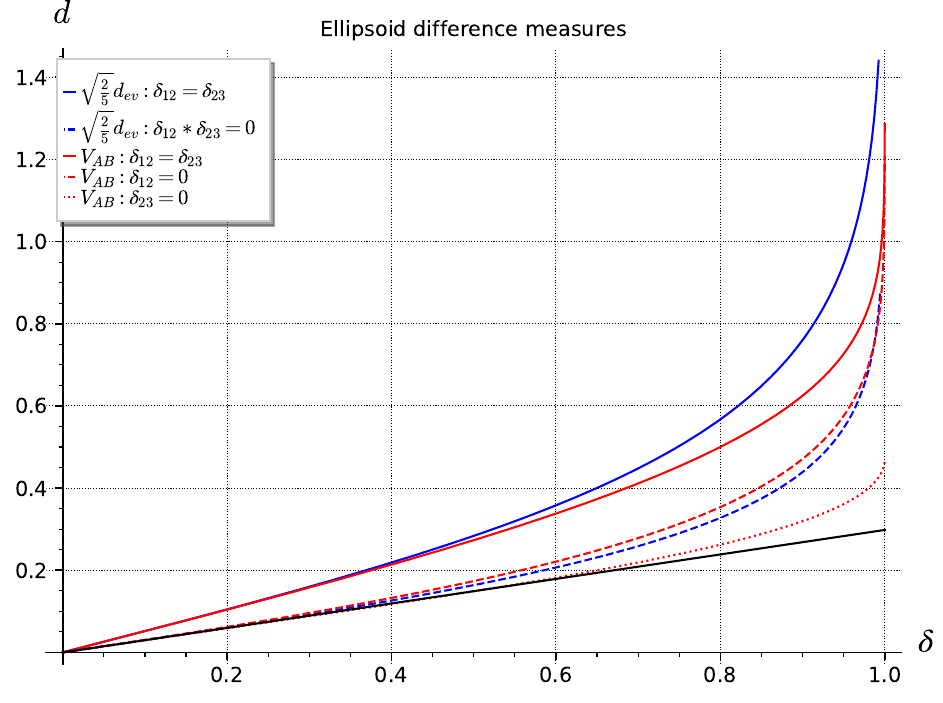}
\par\end{centering}
\caption{Comparison between the continuous measure $V_{AB,\mu}$ (red) and
the eigenvalue measure $d_{ev}$ (blue) for a 3-dimensional ellipsoid
as a function of the eccentricities $\delta_{12}$ and $\delta_{23}$.
Results are shown along the diagonal ($\delta_{12}=\delta_{23}$)
and along the edges $\delta_{12}=0$ (dashed) and $\delta_{23}=0$
(dotted). The eigenvalue measure has been multiplied with $\sqrt{2/5}$
so that the zero limits coincide. On the 2 edges $d_{ev}$ is symmetric,
but the values for $V_{AB,\mu}$ are slightly different.}\label{fig:ellipsoid}
\end{figure}

For the transformation based continuous measure for the 3-dimensional
case we found a similar limit for the 4 measures $V_{AB,\mu}$, $\bar{\gamma}_{\mu}$,
$CV_{\mu}$ and $\mathrm{STRESS_{\mu}}$.
\begin{equation}
\lim_{\delta_{12},\delta_{23}\rightarrow0}V_{AB,\mu}=\frac{2}{3\sqrt{5}}\sqrt{\delta^{2}_{12}+\delta^{2}_{23}+\delta_{12}\delta_{23}}+O\left(\delta^{3}\right)
\end{equation}

with this time a ratio $d_{ev}/V_{AB,\mu}=\sqrt{5/2}$ slightly larger
that the 2d-ratio being equal to $\sqrt{2}$. A comparison between
the continuous measure $V_{AB,\mu}$ and the eigenvalue base measure
$d_{ev}$ is shown in Fig~(\ref{fig:ellipsoid}).

As an example we mention the eigenvalue based $d_{ev,rms}=0.261$
obtained by comparing the MacAdam experimental ellipses with our RieLE
LE \cite{Candry:2022kx} with individual values in the range $d_{ev}=0.082-0.696$.
When comparing the same ellipses with the discrete sampling method
as used by MacAdam \cite{MacAdam:1964hk} we find slightly lower values
$d_{rms}=0.234$ and a range $d=0.071-0.512$. The (Pearson) correlation
between both measures equals $\rho=0.91$.

\section{Pant's geometric measure}

Pant's geometric measure \cite{Raj-Pant:2012kb} is comparable to
our eigenvalue measure since the difference compares ellipses and
not discrete differences. With $A_{\lambda}$ and $A_{\gamma}$ the
areas of the 2 ellipses the complementary measure only depends on
the area of the cross-section
\begin{equation}
\bar{R}=\frac{A_{\lambda}+A_{\gamma}-2A_{\cap}}{A_{\lambda}+A_{\gamma}-A_{\cap}}
\end{equation}

In the degenerate cases where an ellipse is completely contained by
the other ($A_{\gamma}\subset A_{\lambda}$ or $A_{\lambda}\subset A_{\gamma}$)
we get directly
\begin{equation}
\bar{R}=1-\frac{\min\left(\sqrt{\lambda_{1}\lambda_{2}},\sqrt{\gamma_{1}\gamma_{2}}\right)}{\max\left(\sqrt{\lambda_{1}\lambda_{2}},\sqrt{\gamma_{1}\gamma_{2}}\right)}
\end{equation}

To get a scale invariant measure we scale the reference ellipses eigenvalues
with a scale factor $F$ ($\lambda_{i}\rightarrow F\lambda_{i}$).
Then, starting with a small value for $F$ the reference ellipse will
completely encompass the experimental one. Increasing $F$, the ellipses
will eventually touch each other, then cross each other and then again
touch each other but with the reference ellipse now completely within
the experimental ellipse. There is a finite window $F_{\min}\leq F\leq F_{\max}$
within which the ellipses actually cross each other. It is straightforward
to calculate these cross sections and subsequently the areas of the
4 parts constituting the cross-section area $A_{\cap}$ enclosed by
these 4 points. This area can be expressed as a function of 2 angles
\begin{equation}
A_{\cap}=\frac{\phi_{\lambda}}{\sqrt{\lambda_{1}\lambda_{2}}}+\frac{\phi_{\gamma}}{\sqrt{\gamma_{1}\gamma_{2}}}\label{eq:ellipsesareaangles}
\end{equation}

The details for obtaining these angles are given in Appendix~\ref{sec:Pant's-geometric-measure}.

If we apply the affine transformation discussed in §~\ref{sec:Transformation-based-measures}
and apply Pant's difference measure to the transformed ellipses we
must replace $\lambda_{i}\rightarrow1$, $\gamma_{i}\rightarrow\mu_{i}$
and $\Delta\theta$ should be irrelevant. It follows from (\ref{eq:Pant's p})
that $p_{\mu}=q_{\mu}$ but according to (\ref{eq:deltaq}) $q_{\mu}=q_{\lambda}q_{\gamma}\left(1-\delta_{\lambda}\delta_{\gamma}\cos2\Delta\theta\right)$
and again from (\ref{eq:Pant's p}) $q_{\mu}=p$ thus showing that
$p_{\mu}=p$ is invariant for the transformation and this holds also
for $\bar{R}$.

In the limit for small $\delta$'s we find from (\ref{eq:Pant's p})
and (\ref{eq:Rbaropt})
\begin{equation}
\lim_{\delta\rightarrow0}2\left(p-1\right)=\delta^{2}_{\lambda}+\delta^{2}_{\gamma}-2\delta_{\lambda}\delta_{\gamma}\cos2\Delta\theta=\delta^{2}_{\mu}
\end{equation}
\begin{equation}
\lim_{\delta\rightarrow0}\bar{R}_{\mathrm{opt}}=\frac{2}{\pi}\sqrt{2\left(p-1\right)}=\frac{2}{\pi}\delta_{\mu}
\end{equation}

similar to the limits in (\ref{eq:small limit}) and (\ref{eq:dev2dsmall}).
From Fig.\ref{fig:The-difference-measures-2} we note that $\bar{R}$
closely follows $d_{ev}$ and of all measures is the most linear,
with only a slight acceleration with increasing eccentricity $\delta_{\mu}$.

\section{Discussion}

The main properties of the difference measures discussed in this paper
are summarized in Table~\ref{tab:Overview-of-the}. We started by
considering the discrete measures $V_{AB}$, $\bar{\gamma}$, $CV$
, which are often combined into a single measure $\mathrm{PF}/3$
, and $\mathrm{STRESS}$. Considering a rather extreme case where
only 2 directions of an ellipse are sampled, we could find no need
for taking the average $\mathrm{PF/3}$ since the variation of $V_{AB}$,
$\bar{\gamma}$ and $CV$ around the ellipse are very similar, with
exactly the same nodes (see Fig.\ref{fig:Discrete-difference-measures}).
It is obvious that properly sampling the directions in a color point
is paramount and inadequate sampling is very likely the cause of possible
discrepancies observed between difference measures. To eliminate this
sampling error we defined continuous measures by integrating over
all directions. These 4 measures then have exactly the same behavior
for small deviations from circularity and differences are only visible
for large deviations from circularity but are qualitative still similar.
Considering again a rather extreme case by comparing 2 perpendicular
ellipses with the same aspect ratio, $\bar{\gamma}$ gives the largest
result, followed by $V_{AB}$, $CV$ and $\mathrm{STRESS}$ with the
latter 3 close together (see Fig.\ref{fig:The-difference-measures}).
We conclude that $\bar{\gamma}$ penalizes large errors more heavily
and $\mathrm{STRESS}$ the least.

These 4 basic measures are scale invariant and symmetric, except for
$CV$ which is not symmetric. Considering its close relation with
$\mathrm{STRESS}$ we can drop it from further consideration. Only
one of the remaining measures is also coordinate independent ($\mathrm{STRESS}$).
Exploiting the coordinate independence we affinely transform both
ellipses in the same way and so that one of them becomes a unit ball.
Comparing these transformed ellipses is much easier, since the number
of independent variables is reduced. Compared with the original results,
the transformed results, labeled as $\bar{\gamma}_{\mu}$, $V_{AB,\mu}$,
${CV}_{\mu}$ are not exactly the same (see Fig.\ref{fig:coordinateIndependence})
but the changes are still modest. These transformation based measures
are functions of a single parameter $\delta_{\mu}=\frac{\mu_{1}}{\mu_{2}}$
and the behavior is largely the same as for the general continuous
measures (see Fig.\ref{fig:The-difference-measures-2}): $\bar{\gamma}_{\mu}$
is still on top, now followed by ${CV}_{\mu}$ and it turns out that
$V_{AB,\mu}\approx\mathrm{STRESS}$. Considering this coincidence
and the fact that $\mathrm{STRESS}$ is coordinate independent we
can drop $V_{AB,\mu}$ too.

Table.\ref{tab:Overview-of-the} contains also an outlier, namely
the $r$ measure, a classic correlation coefficient which may perhaps
make statistical sense \cite{Kirchner:2011zy} but contrary to all
the other measures it heavily penalizes misalignment (see Fig.\ref{fig:The-continuous-correlation}).
Although this might be considered as a benefit this is accompanied
by a severe dependency on the angle $\Delta\theta$ also in situations
where it has no physical meaning.

\begin{table}
\caption{Overview of the difference measures considered. The $\gamma,r$ and
Pant's geometric $R$ have been replaced by their complements $\bar{\gamma}=\gamma-1$,
$\bar{r}=1-r$ and $\bar{R}=1-R$ respectively. Pant's measure has
also been made scale invariant. The limits in the last columns are
expressed in terms of $\delta_{\mu}\approx\sqrt{\delta^{2}_{\lambda}+\delta^{2}_{\gamma}-2\delta_{\lambda}\delta_{\gamma}\cos2\Delta\theta}$.
The next to last column indicates whether the globally optimized difference
is easily obtainable from the locally optimized ones.}\label{tab:Overview-of-the}

\centering{}%
\begin{tabular}{cccccc}
\toprule 
 & \begin{cellvarwidth}[t]
\centering
coordinate \\
independent
\end{cellvarwidth} & symmetric & \begin{cellvarwidth}[t]
\centering
scale\\
invariant
\end{cellvarwidth} & \begin{cellvarwidth}[t]
\centering
local/global\\
relation
\end{cellvarwidth} & \begin{cellvarwidth}[t]
\centering
$\lim_{\delta\rightarrow0}$\\
(chromatic)
\end{cellvarwidth}\tabularnewline
\midrule
\midrule 
$V_{AB}$ & \xmark & $\checkmark$ & $\checkmark$ & \xmark & $\frac{1}{2\sqrt{2}}\delta_{\mu}$\tabularnewline
\midrule 
$\bar{\gamma}$ & \xmark & $\checkmark$ & $\checkmark$ & $\checkmark$ & $\frac{1}{2\sqrt{2}}\delta_{\mu}$\tabularnewline
\midrule
\midrule 
$\mathrm{STRESS}$ & $\checkmark$ & $\checkmark$ & $\checkmark$ & \xmark & $\frac{1}{2\sqrt{2}}\delta_{\mu}$\tabularnewline
\midrule 
$CV$ & \xmark & {\xmark} & $\checkmark$ & \xmark & $\frac{1}{2\sqrt{2}}\delta_{\mu}$\tabularnewline
\midrule
\midrule 
$\bar{r}$ & \xmark & $\checkmark$ & $\checkmark$ & \xmark & $1-\cos2\Delta\theta$\tabularnewline
\midrule
\midrule 
$d_{\mathrm{ev}}=\bar{\gamma}_{ev}$ & $\checkmark$ & $\checkmark$ & $\checkmark$ & $\checkmark$ & $\frac{1}{2}\delta_{\mu}$\tabularnewline
\midrule 
$\bar{R}$ & $\checkmark$ & $\checkmark$ & $\checkmark$ & \xmark & $\frac{2}{\pi}\delta_{\mu}$\tabularnewline
\bottomrule
\end{tabular}
\end{table}

After applying the affine transformation the different criteria merely
measure the deviation of an ellipsoid/ellipse from the unit ball.
There is no real need to sample the directions or integrate over all
of them. The deviation can equally well and much more easily be obtained
using just the 2 or 3 eigenvalues and this brings us with our proposed
$d_{ev}$, which with hindsight could also have been labeled as $\bar{\gamma}_{ev}$.
It must be emphasized that all eigenvalue based measures are by definition
coordinate independent, whereas $\bar{\gamma}$ itself is not. With
${CV}$ and $V_{AB}$ dropped earlier it would still make sense to
consider an eigenvalue based $\mathrm{STRESS}_{ev}$. However it turns
out that the relation between the locally (in a single color point)
optimized ellipsoids and the globally optimized ones is much simpler
for the $\bar{\gamma}$ measure. Suppose we found the optimal scaling
factors and difference measures in a given set of color points (Here
we use the double indices notation with index $k$ labeling the color
points and index $l$ the directions in a color point.) according
to (\ref{eq:gammaFopt}) and (\ref{eq:gamma})
\begin{equation}
\ln F_{k}=-\left\langle \ln\mathrm{\left(E/V\right)}_{l}\right\rangle _{k}\label{eq:lnFk}
\end{equation}
\begin{equation}
\left(\ln\gamma_{\mathrm{opt},k}\right)^{2}=\left\langle \left(\ln\mathrm{\left(E/V\right)}_{l}\right)^{2}\right\rangle _{k}-\left(\ln F_{k}\right)^{2}\label{eq:gammaoptk}
\end{equation}

the average being taken over all directions $l$ in a color point
$k$. The overall optimal scaling factor can be found in the same
way but by averaging over all data
\begin{equation}
\ln F_{\mathrm{opt}}=-\overline{\left\langle \ln\mathrm{\left(E/V\right)}_{l}\right\rangle _{k}}=\overline{\ln F_{k}}
\end{equation}

where we denote the averaging over the color points by an overline
and used (\ref{eq:lnFk}) in the 2nd step. Likewise the resulting
optimal difference is given by
\begin{equation}
\left(\ln\gamma_{\mathrm{opt}}\right)^{2}=\overline{\left\langle \left(\ln\mathrm{\left(E/V\right)}_{l}\right)^{2}\right\rangle _{k}}-\left(\ln F_{opt}\right)^{2}
\end{equation}

Using (\ref{eq:gammaoptk}) this can be written as
\begin{equation}
\left(\ln\gamma_{\mathrm{opt}}\right)^{2}=\overline{\left(\ln\gamma_{\mathrm{opt},k}\right)^{2}}+\overline{\left(\ln F_{k}\right)^{2}}-\left(\overline{\ln F_{k}}\right)^{2}
\end{equation}

The global optimal difference is thus found by taking the average
of the locally defined optimal differences augmented by the variance
of the scaling factors. A similar reasoning can be followed for e.g.
$\mathrm{STRESS}$ but due to the occurrence of additional averages
this becomes more complicated and one needs to resort to weighted
averages, the weights being based on $\left\langle \Delta E^{2}_{l}\right\rangle _{k}$
and $\left\langle \Delta V^{2}_{l}\right\rangle _{k}$. Finally we
mention that the scaling factor $F$ has been applied to $\Delta E$
(see (\ref{eq:gammabase})) and thus to $\left\Vert x\right\Vert _{V}=\left\Vert x\right\Vert _{\Delta V=1}$
and therefore $F=d\sigma$ of the experimental ellipsoid.

Pant's geometrically defined difference measure has many interesting
properties, being symmetric and coordinate independent and is easily
extended to become scale invariant but as for the $\mathrm{STRESS}$
the relation between the globally defined scaling factor and the locally
defined ones is likely problematic. For the locally optimized ellipses
the degenerated condition does not occur and the optimal scaling factor
is always contained within its bounds. This cannot longer be guaranteed
for the global optimization and this will complicate the relation.

Taking all considerations into account our eigenvalue based measure,
comparable to the discrete $\bar{\gamma}$ measure, is the only one
meeting all desirable properties listed in Table~\ref{tab:Overview-of-the},
combining in particular the coordinate independence of STRESS and
the variance additivity of $\bar{\gamma}$. Since $d^{2}_{ev}$ is
nothing more than a standard variance, it can also be used to perform
F-tests, comparable to STRESS. If the data do not allow the proper
definition of an ellipsoid and one must rely on a discrete method
the choice should be made between $\bar{\gamma}$, which is the most
sensitive but coordinate dependent and STRESS which is coordinate
independent, but the least sensitive.

\subsection*{Acknowledgment}

We are much indebted to Kristiaan Neyts for suggesting the use of
Eq.\ref{eq:d-base}.

\bibliographystyle{unsrturl}
\bibliography{referencesrev2.bib}

\appendix

\section{Elliptic Integrals \cite{Carlson:2025gx}.}\label{sec:Elliptic-Integrals.}

For convenience we list the elliptic integrals used in the main text,
with their standard Legendre forms and their more modern symmetric
forms. For the Legendre forms we use the so-called parameter $m=k^{2}$,
where $k$ is the modulus, as the argument and we only have to deal
with complete integrals.

Complete elliptic integral of the first kind
\begin{equation}
K\left(m\right)=\int^{\pi/2}_{0}\frac{d\theta}{\sqrt{1-m\sin^{2}\theta}}
\end{equation}

Complete elliptic integral of the second kind
\begin{equation}
E\left(m\right)=\int^{\pi/2}_{0}\sqrt{1-m\sin^{2}\theta}d\theta
\end{equation}

Complete elliptic integral of the third kind
\begin{equation}
\Pi\left(n,m\right)=\int^{\pi/2}_{0}\frac{d\theta}{\left(1-n\sin^{2}\theta\right)\sqrt{1-m\sin^{2}\theta}}
\end{equation}

Symmetric elliptic integral of the first kind
\begin{equation}
R_{F}\left(x,y,z\right)=\frac{1}{2}\int^{\infty}_{0}\frac{dt}{s\left(t\right)}
\end{equation}

where
\begin{equation}
s\left(t\right)=\sqrt{\left(t+x\right)\left(t+y\right)\left(t+z\right)}
\end{equation}

Symmetric elliptic integral of the second kind
\begin{equation}
R_{G}\left(x,y,z\right)=\frac{1}{4}\int^{\infty}_{0}\frac{1}{s\left(t\right)}\left[\frac{x}{t+x}+\frac{y}{t+y}+\frac{z}{t+z}\right]tdt
\end{equation}

Symmetric elliptic integral of the third kind
\begin{equation}
R_{J}\left(x,y,z,p\right)=\frac{3}{2}\int^{\infty}_{0}\frac{dt}{\left(t+p\right)s\left(t\right)}
\end{equation}

These symmetric integrals are complete if exactly one of the arguments
$x,y,z$ is zero. An additional function is defined by
\[
R_{D}\left(x,y,z\right)=R_{J}\left(x,y,z,z\right)
\]

and a useful connection formula is given by
\begin{equation}
\frac{p}{3}R_{J}\left(0,y,z,p\right)+\frac{q}{3}R_{J}\left(0,y,z,q\right)=R_{F}\left(0,y,z\right)\quad pq=yz\label{eq:RJconnection}
\end{equation}

Many relations exist between the Legendre and the symmetric forms
with the main ones
\begin{equation}
K\left(1-m\right)=R_{F}\left(0,m,1\right)
\end{equation}
\begin{equation}
E\left(1-m\right)=2R_{G}\left(0,m,1\right)
\end{equation}
\begin{equation}
\Pi\left(n,1-m\right)-K\left(1-m\right)=\frac{1}{3}nR_{J}\left(0,m,1,1-n\right)
\end{equation}
\begin{equation}
K\left(1-m\right)-E\left(1-m\right)=\frac{1}{3}\left(1-m\right)R_{D}\left(0,m,1\right)
\end{equation}
\begin{equation}
E\left(1-m\right)-mK\left(1-m\right)=\frac{1}{3}m\left(1-m\right)R_{D}\left(0,1,m\right)\label{eq:con5}
\end{equation}

\section{Detailed calculation of eq.(\ref{eq:I0def}) and eq.(\ref{eq:I2def})}\label{sec:Detailed-calculation-of}

The parameters occurring in the integrals $I_{0}$ and $I_{2}$ defined
in (\ref{eq:I0def}) and (\ref{eq:I2def}) are given by
\begin{equation}
a^{2}=\frac{g_{11}}{g_{22}}=\frac{1-\delta_{\gamma}\cos2\Delta\theta}{1+\delta_{\gamma}\cos2\Delta\theta}\label{eq:a2def}
\end{equation}
\begin{equation}
b=-\frac{g_{12}}{g_{22}}=\frac{\delta_{\gamma}\sin2\Delta\theta}{1+\delta_{\gamma}\cos2\Delta\theta}\label{eq:bdef}
\end{equation}
with $a^{2}\geq b^{2}$ and
\begin{equation}
c^{2}=\frac{g^{\mathrm{ref}}_{11}}{g^{\mathrm{ref}}_{22}}=\frac{1-\delta_{\lambda}}{1+\delta_{\lambda}}=\frac{\lambda_{1}}{\lambda_{2}}\label{eq:c2def}
\end{equation}

The integral $I_{0}$ is an elliptic integral which can be found using
standard procedures \cite{Carlson:2025gx}, in particular by a change
of variables 
\begin{equation}
u=\frac{t-\nu^{\prime}_{1}b}{t+\nu^{\prime}_{2}b}
\end{equation}

where $\nu^{\prime}_{1}=\frac{\nu_{1}}{\nu_{1}-1}$, $\nu^{\prime}_{2}=\frac{\nu_{2}}{1-\nu_{2}}$
with $\nu_{1,2}$ the roots of a quadratic with coefficients depending
on $a,b,c$
\begin{equation}
\nu_{1,2}\left(a^{2}-b^{2}\right)=\frac{a^{2}+c^{2}}{2}\pm\sqrt{\left(\frac{a^{2}-c^{2}}{2}\right)^{2}+b^{2}c^{2}}\label{eq:nu12def}
\end{equation}
The roots are real and positive with $\nu_{1}\geq1$, $\nu_{2}\leq1$,
$\nu_{1}+\nu_{2}>1$. Then
\begin{equation}
I_{0}\left(\delta_{\lambda},\delta_{\gamma},\Delta\theta\right)=\frac{2/\pi}{\sqrt{\nu_{1}\left(a^{2}-b^{2}\right)}}R_{F}\left(0,\frac{\nu_{2}}{\nu_{1}},1\right)\label{eq:I0result}
\end{equation}

In the limit $b\rightarrow0$ integral $I_{0}$ becomes symmetric
in $a,c$ with $\frac{\nu_{2}}{\nu_{1}}=\frac{\min\left(a^{2},c^{2}\right)}{\max\left(a^{2},c^{2}\right)}$
and $v_{1}a^{2}=\max\left(a^{2},c^{2}\right)$. In that case no change
of variables is required and the integral is readily reduced to a
symmetric elliptic integral.

Using the definitions (\ref{eq:a2def})-(\ref{eq:c2def}) we find
\begin{equation}
\frac{a^{2}+c^{2}}{2}=\frac{1-\delta_{\lambda}\delta_{\gamma}\cos2\Delta\theta}{N}
\end{equation}
\begin{equation}
\frac{a^{2}-c^{2}}{2}=\frac{\delta_{\lambda}-\delta_{\gamma}\cos2\Delta\theta}{N}
\end{equation}

where the denominator $N=\left(1+\delta_{\lambda}\right)\left(1+\delta_{\gamma}\cos2\Delta\theta\right)$.
Substituting these in (\ref{eq:nu12def}) we get
\begin{equation}
N\nu_{1,2}\left(a^{2}-b^{2}\right)=1-\delta_{\lambda}\delta_{\gamma}\cos2\Delta\theta\pm\sqrt{D}\label{eq:}
\end{equation}

where $D=\delta^{2}_{\lambda}+\delta^{2}_{\gamma}-2\delta_{\lambda}\delta_{\gamma}\cos2\Delta\theta-\delta^{2}_{\lambda}\delta^{2}_{\gamma}\sin^{2}2\Delta\theta$,
and from which it readily follows that
\begin{equation}
\sqrt{\frac{\nu_{2}}{\nu_{1}}}=\frac{\sqrt{1-\delta^{2}_{\lambda}}\sqrt{1-\delta^{2}_{\gamma}}}{1-\delta_{\lambda}\delta_{\gamma}\cos2\Delta\theta+\sqrt{D}}=\frac{1-\delta_{\lambda}\delta_{\gamma}\cos2\Delta\theta-\sqrt{D}}{\sqrt{1-\delta^{2}_{\lambda}}\sqrt{1-\delta^{2}_{\gamma}}}\label{eq:nu2onu1result}
\end{equation}

and whence
\begin{equation}
\nu_{1}\left(a^{2}-b^{2}\right)=\sqrt{\frac{\nu_{1}}{\nu_{2}}}\frac{\sqrt{1-\delta^{2}_{\lambda}}\sqrt{1-\delta^{2}_{\gamma}}}{\left(1+\delta_{\lambda}\right)\left(1+\delta_{\gamma}\cos2\Delta\theta\right)}\label{eq:nu1result}
\end{equation}

Reflecting one ellipse w.r.t. the main axes of the other leaves all
averages invariant. Therefore $\nu_{i}$ and $\nu^{\prime}_{i}$ remain
invariant when $\Delta\theta$ and $b$ change sign.

Integral $I_{2}$ can be found using the same procedure but due to
the additional factor $1+t^{2}$ in the denominator gives rise to
an additional polynomial in $u$ 
\begin{equation}
P\left(u\right)=\left(1+\nu^{\prime2}_{2}b^{2}\right)\left(u-z\right)\left(u-z^{*}\right)
\end{equation}
with complex roots $z$ and $z^{*}$ where 
\begin{equation}
z=\frac{-b\nu^{\prime}_{1}+i}{b\nu^{\prime}_{2}+i}
\end{equation}
\begin{equation}
z-z^{*}=\frac{2ib\left(\nu^{\prime}_{1}+\nu^{\prime}_{2}\right)}{1+\nu^{\prime2}_{2}b^{2}}
\end{equation}
\begin{equation}
\frac{z+z^{*}}{2}=\frac{1-\nu^{\prime}_{2}\nu^{\prime}_{1}b^{2}}{1+\nu^{\prime2}_{2}b^{2}}
\end{equation}
\begin{equation}
zz^{*}=\frac{1+\nu^{\prime2}_{1}b^{2}}{1+\nu^{\prime2}_{2}b^{2}}
\end{equation}

This polynomial can be handled by fractional decomposition and eventually
leads to the result
\begin{multline}
I_{2}=\frac{2/\pi}{\sqrt{\nu_{1}\left(a^{2}-b^{2}\right)}}\left\{ \frac{1}{1+\nu^{\prime2}_{2}b^{2}}R_{F}\left(0,\frac{\nu_{2}}{\nu_{1}},1\right)\right.\\
-\frac{1}{3}b\left(\nu^{\prime}_{1}+\nu^{\prime}_{2}\right)\frac{\nu^{\prime}_{2}}{\nu^{\prime}_{1}}\left.\Im\left[\frac{b\nu^{\prime}_{1}-i}{\left(b\nu^{\prime}_{2}+i\right)^{3}}R_{J}\left(0,\frac{\nu_{2}}{\nu_{1}},1,-\frac{\nu^{\prime}_{2}}{\nu^{\prime}_{1}}\left(\frac{b\nu^{\prime}_{1}-i}{b\nu^{\prime}_{2}+i}\right)^{2}\right)\right]\right\} \label{eq:I2result}
\end{multline}

where in general the last argument of $R_{J}\left(x,y,z,p\right)$
is complex. These function evaluations have been implemented following
\cite{Carlson:1995qu}.

If the ellipses are aligned ($\Delta\theta=0$) or perpendicular ($\Delta\theta=\frac{\pi}{2}$)
then $b=0$ and again no change of variables is needed and integral
$I_{2}$ is then also symmetric in $a,c$. These limiting integrals
can be found more easily as a combination of complete symmetric elliptic
integrals with real arguments, but eq.(\ref{eq:I2result}) also reduces
to this solution for $b\rightarrow0$.

If $a>c$ then $\nu_{1}\rightarrow1+\frac{b^{2}}{a^{2}-c^{2}}\rightarrow1$,
$\nu^{\prime}_{1}\rightarrow\frac{a^{2}-c^{2}}{b^{2}}\rightarrow\infty$,
$\nu_{2}\rightarrow\frac{c^{2}}{a^{2}}$, $\nu^{\prime}_{2}\rightarrow\frac{c^{2}}{a^{2}-c^{2}}$,
$-\frac{\nu^{\prime}_{2}}{\nu^{\prime}_{1}}\left(\frac{b\nu^{\prime}_{1}-i}{b\nu^{\prime}_{2}+i}\right)^{2}\rightarrow c^{2}$
and $b\left(\nu^{\prime}_{1}+\nu^{\prime}_{2}\right)\frac{\nu^{\prime}_{2}}{\nu^{\prime}_{1}}\frac{b\nu^{\prime}_{1}-i}{\left(b\nu^{\prime}_{2}+i\right)^{3}}\rightarrow ic^{2}$
turning (\ref{eq:I2result}) into
\begin{equation}
I_{2}\left(b=0,a>c\right)=\frac{2/\pi}{a\sqrt{\nu_{1}}}\left\{ R_{F}\left(0,\frac{\nu_{2}}{\nu_{1}},1\right)-\frac{c^{2}}{3}R_{J}\left(0,\frac{\nu_{2}}{\nu_{1}},1,c^{2}\right)\right\} \label{eq:I2b0agtc}
\end{equation}

If $a<c$ then $\nu_{1}\rightarrow\frac{c^{2}}{a^{2}}$, $\nu^{\prime}_{1}\rightarrow\frac{c^{2}}{c^{2}-a^{2}}$,
$\nu_{2}\rightarrow1-\frac{b^{2}}{c^{2}-a^{2}}$, $\nu^{\prime}_{2}\rightarrow\frac{c^{2}-a^{2}}{b^{2}}\rightarrow\infty$,
$-\frac{\nu^{\prime}_{2}}{\nu^{\prime}_{1}}\left(\frac{b\nu^{\prime}_{1}-i}{b\nu^{\prime}_{2}+i}\right)^{2}\rightarrow\frac{1}{c^{2}}$
and $b\left(\nu^{\prime}_{1}+\nu^{\prime}_{2}\right)\frac{\nu^{\prime}_{2}}{\nu^{\prime}_{1}}\frac{b\nu^{\prime}_{1}-i}{\left(b\nu^{\prime}_{2}+i\right)^{3}}\rightarrow-i\frac{1}{c^{2}}$
turning (\ref{eq:I2result}) into
\begin{align}
I_{2}\left(b=0,a<c\right) & =\frac{2/\pi}{a\sqrt{\nu_{1}}}\frac{1}{3c^{2}}R_{J}\left(0,\frac{\nu_{2}}{\nu_{1}},1,\frac{1}{c^{2}}\right)\label{eq:I2b0altc}\\
 & =\frac{2/\pi}{a\sqrt{\nu_{1}}}\left\{ R_{F}\left(0,\frac{\nu_{2}}{\nu_{1}},1\right)-\frac{a^{2}}{3}R_{J}\left(0,\frac{\nu_{2}}{\nu_{1}},1,a^{2}\right)\right\} 
\end{align}

where in both cases $\frac{\nu_{2}}{\nu_{1}}=\frac{\min\left(a^{2},c^{2}\right)}{\max\left(a^{2},c^{2}\right)}$
and $\nu_{1}=\max\left(1,\frac{c^{2}}{a^{2}}\right)$. The expression
on the 2nd line follows from (\ref{eq:RJconnection}) (with $pq=\frac{\nu_{2}}{\nu_{1}}=\frac{a^{2}}{c^{2}}$)
and is symmetric with (\ref{eq:I2b0agtc}).

\section{Pant's geometric measure}\label{sec:Pant's-geometric-measure}

Consider as in the main text a reference ellipse with eigenvalues
$\lambda_{1}\leq\lambda_{2}$, aligned with the axes $\left(x^{1},x^{2}\right)$
and a second ellipse with eigenvalues $\gamma_{1}\leq\gamma_{2}$
whose long axis makes an angle $\Delta\theta$ with the $x^{1}$-axis.
We assume the ellipse boundaries actually cross each other, the case
where one ellipse is completely contained within the other being straightforward
to handle, then Pant's (complementary) difference measure $\bar{R}$
(but extended by inclusion of a scaling factor $F$ of the eigenvalues
of the reference ellipse) depends, besides on the scaling factor,
on a single parameter $p$, defined by
\begin{equation}
p=q_{\lambda}q_{\gamma}\left(1-\delta_{\lambda}\delta_{\gamma}\cos2\Delta\theta\right)\geq1\label{eq:Pant's p}
\end{equation}

Defining an auxiliary quantity
\begin{equation}
D_{p}\left(x\right)=2\left(p-1\right)\left(1-x\right)-x^{2}
\end{equation}
where $1-x=\frac{F}{\sqrt{\mu_{1}\mu_{2}}}$ contains the scaling
factor, the area of the cross-section of these ellipses can be expressed
in terms of 2 equivalent angles (see (\ref{eq:ellipsesareaangles}))
and these are given by
\begin{equation}
\tan\phi_{\lambda}\left(x\right)=\frac{\sqrt{D_{p}}}{p-1+x}\quad\tan\phi_{\lambda}\left(x\right)=\frac{\sqrt{D_{p}}}{p\left(1-x\right)-1}
\end{equation}
where proper 2-argument functions should be used for the inverses
\begin{equation}
\phi_{\lambda}\left(x\right)=\arctan\left(p-1+x,\sqrt{D_{p}}\right)\quad\phi_{\lambda}\left(x\right)=\arctan\left(p\left(1-x\right)-1,\sqrt{D_{p}}\right)
\end{equation}

The difference measure follows as
\begin{equation}
\bar{R}\left(x\right)=2\frac{\pi-\left(\phi_{\lambda}+\phi_{\gamma}\right)-x\left(\frac{\pi}{2}-\phi_{\gamma}\right)}{2\pi-\left(\phi_{\lambda}+\phi_{\gamma}\right)-x\left(\pi-\phi_{\gamma}\right)}
\end{equation}

The scaling parameter $x$ is bounded by $x_{\min}=p-\sqrt{p^{2}-1}$
and $x_{\max}=p+\sqrt{p^{2}-1}$ and the optimum is obtained for $x_{\mathrm{opt}}=0$,
thus $F_{opt}=\sqrt{\mu_{1}\mu_{2}}$, with
\begin{equation}
D_{p}\left(0\right)=2\left(p-1\right)
\end{equation}
\begin{equation}
\phi_{\lambda}\left(0\right)=\phi_{\gamma}\left(0\right)=\arctan\left(p-1,\sqrt{2\left(p-1\right)}\right)
\end{equation}
 {}
\begin{equation}
\bar{R}_{\mathrm{opt}}=\frac{1-\frac{2}{\pi}\phi_{\lambda/\gamma}\left(0\right)}{1-\frac{1}{\pi}\phi_{\lambda/\gamma}\left(0\right)}\label{eq:Rbaropt}
\end{equation}

\end{document}